\documentclass[a4paper,twocolumn,11pt,unpublished]{quantumarticle}
\pdfoutput=1
\usepackage[utf8]{inputenc}
\usepackage[english]{babel}
\usepackage[T1]{fontenc}
\usepackage{amsmath}
\usepackage{hyperref}
\usepackage{tikz}
\usepackage{lipsum}
\usepackage[numbers,sort&compress]{natbib}
\usepackage{graphicx, color,ulem,pdfcomment,lipsum,capt-of,amsmath,booktabs}

\begin{document}

\title{Interaction-induced wavefunction collapse}

\author{Arnab Acharya}
\affiliation{Indian Institute Science Education and Research Kolkata, West Bengal, India}
\orcid{0000-0002-4711-4262}

\author{Pratik Jeware}

\author{Soumitro Banerjee}
\email{soumitro@iiserkol.ac.in}
\orcid{0000-0003-3576-0846}
\affiliation{Indian Institute Science Education and Research Kolkata, West Bengal, India}

\maketitle

\begin{abstract}
    Almost a century after the development of quantum mechanics, we still do not have a consensus on the process of collapse of wavefunctions. Some theories require the intervention of a conscious observer while some see it as a stochastic process, and most theories violate energy conservation. In this paper we hypothesise that the collapse of wavefunctions can be caused by interactions with other objects (macroscopic or microscopic) and energy is conserved in that process. To test various hypotheses regarding collapse of wavefunctions, we propose a model system which is the quantum analogue of a classical soft-impact oscillator. We propose some alternative postulates regarding the conditions for and the result of a collapse, and obtain the implication of each on the behavior of observable quantities, which can possibly be experimentally tested.
\end{abstract}

\section{Introduction}

 Even though the theoretical structure of quantum mechanics is almost a century old, we have very little understanding of an important component of that theoretical structure---the collapse of the wavefunction. This concept is invoked to explain the results of experiments like the double-slit experiment. But it is largely ignored (or not needed) in most successful applications of quantum mechanics \cite{griffiths_schroeter_2018,sep-qm-collapse}.

 The mainstream Copenhagen interpretation of quantum mechanics claims that, whenever an observation is made, the wavefunction of a quantum system collapses instantly to an eigenstate of the observable being measured. Many questions have been raised \cite{dewitt2015quantum} regarding what constitutes an observation.  ``It would seem that the theory is exclusively concerned about `results of measurement', and has nothing to say about anything else'' wrote John Bell \cite{bell1990against}. ``What exactly qualifies some physical systems to play the role of `measurer'? Was the wavefunction of the world waiting to jump for thousands of millions of years until a single-celled living creature appeared? Or did it have to wait a little longer, for some better qualified system $\cdots$ with a PhD?''  These questions have remained largely unanswered, and are considered to belong to the domain of philosophy rather than physics.

 The founders of quantum mechanics did not provide an explicit mechanism for the collapse of the wavefunction and later it was termed as the measurement problem. Einstein believed that a complete description of physical reality would not be possible in this theory \cite{dolling2003tests}. The famous Einstein-Podolsky-Rosen \textit{gedankenexperiment} \cite{einstein1935can} illuminated the non-local structure of quantum entanglement and used it as an objection against the completeness of quantum mechanics. Einstein's convictions about the determinism of nature inspired hidden variable theories which tried to rid quantum mechanics of indeterminacy. John Bell \cite{bell1964einstein} showed that local hidden variable theories are incompatible with the predictions of quantum mechanics. Since then there have been several attempts to develop theories that incorporate a mechanism for wavefunction collapse.

Some have advocated for decoherence \cite{joos2013decoherence,schlosshauer2007decoherence} as a means to solve the measurement problem \cite{pessoa1997can}. Even though it explains the absence of macroscopic superposition and the emergence of the classical world, it remains doubtful  whether it has solved the measurement problem---as the founders of decoherence theory admit in their seminal papers \cite{adler2003decoherence,joos1985emergence}.

\textit{Dynamical collapse theories} \cite{bassi2013models}, on the other hand, supplant the unitary evolution with stochasticity and nonlinearity in such a way that all predictions of quantum mechanics are approximately reproduced in the microscopic limit while precluding macroscopic superpositions like Schr\"odinger cat states. A variety of collapse models have been developed that differ in their localization basis: while \cite{milburn1991intrinsic,benatti1988operations} and \cite{bassi2004numerical} use the energy, momentum and spin basis respectively, the Ghirardi-Rimini-Weber (GRW) \cite{ghirardi1986unified} and the \textit{Continuous Spontaneous Localisation} (CSL)\cite{ghirardi1990markov} operate in the position basis. The GRW model postulates wavefunction collapse as a random and spontaneous process. The collapsed wavefunction has a Gaussian form which is in turn dependent on certain natural constants that the theory introduces. Similar to it is the CSL model, the only difference being that the wavefunction in CSL collapses continuously in time \cite{pearle1989combining}. This was followed by the Diosi-Penrose model \cite{penrose1996gravity} which is based on \textit{The Quantum Mechanics With Universal Position Localisation} (QMUPL) \cite{diosi1989models} where gravity is responsible for the collapse of a wavefunction. These models are also characterized by the noise they use for generating the stochasticity they need to exhibit collapse.

John Bell had argued in favour of seeing the collapse of wavefunction as a natural process: ``If the theory is to apply to anything but highly idealised laboratory operations, are we not obliged to admit that more or less `measurement-like' processes are going on more or less all the time, more or less everywhere? Do we not have jumping then all the time?'' \cite{bell1990against}
 
In this paper, we hypothesise that collapse of a wavefunction does not require the intervention of a conscious observer. An interaction with another physical object---macroscopic or microscopic---can also cause the collapse of a wavefunction in the position space. The probability of occurrence of such an interaction depends on the overlap between their probability density functions.

 If collapse-like processes indeed abound, as they reasonably should, the violation in energy conservation should be quite observable. The absence of such observations makes us hypothesise that any valid collapse mechanism must satisfy energy conservation, at least on average. 

 In order for a collapse model to be admissible, one has to propose postulates which have experimentally testable predictions. In this paper, we propose a model which can be used to test such postulates about interaction-induced wavefunction collapse, and work out the predictions of the various possible postulates when applied to that
 system.
 
\section{The model system}
\begin{figure} 
  \centering
  \includegraphics[trim={7cm 5.5cm 7cm 8.35cm},clip,width=0.7\linewidth]{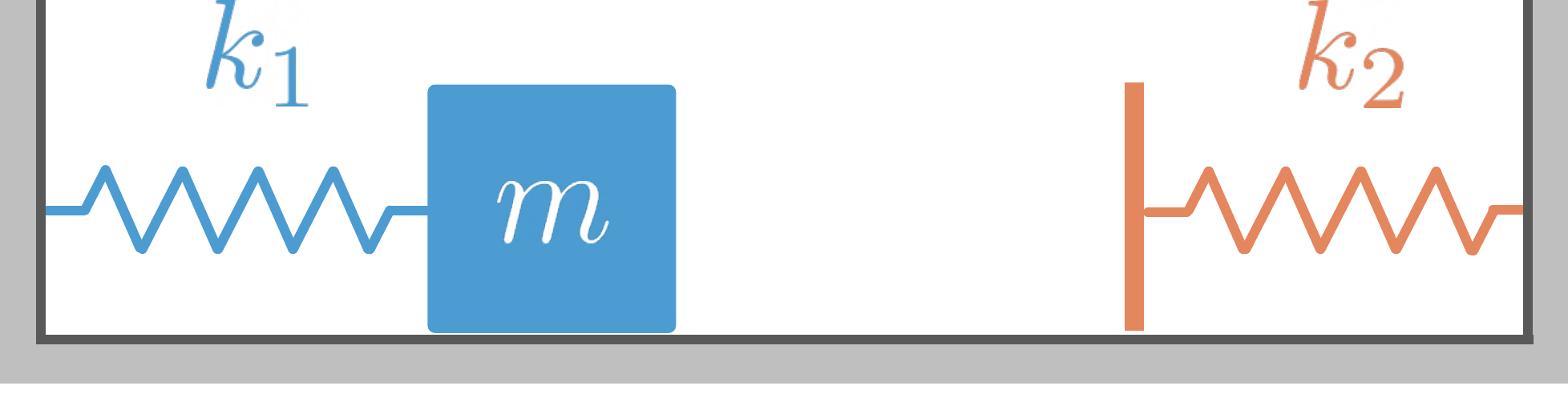}
  \caption{The classical soft-impact oscillator\label{softimpact}.}
\end{figure}

Fig.~\ref{softimpact} shows the classical analog of the system under consideration---a simple harmonic oscillator with mass $m$ and spring constant $k_1$ which can impact with a massless wall, cushioned by a spring of constant $k_2$. The variable $x$ is measured from the equilibrium position of the mass, and the wall is at $x_{\rm wall}$ when the spring $k_2$ is relaxed. 

The classical system, with the inclusion of damping and external
forcing, is known to exhibit a rich variety of dynamical phenomena
which are initiated when the mass grazes the wall
\cite{peterka2004phenomena,ing2008experimental}. 
  We choose this particular system as the wall may be modeled either
  classically or quantum mechanically. This allows us to investigate
  the collapse of the wavefunction due to interaction with either a
  macroscopic or microscopic system.



\begin{figure*}
\centering
\includegraphics[width=0.235\linewidth]{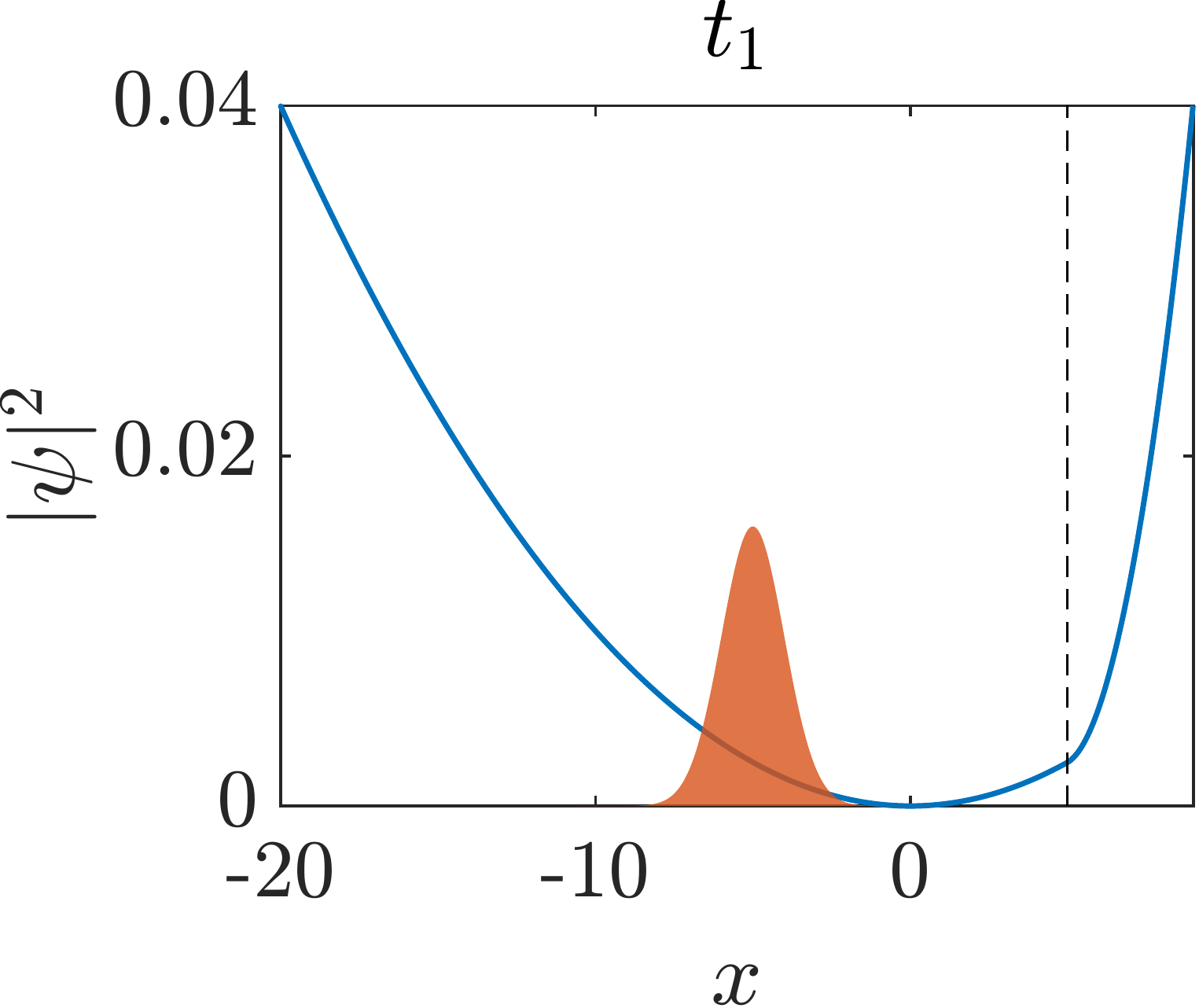}\hspace{0.01\linewidth}
\includegraphics[width=0.235\linewidth]{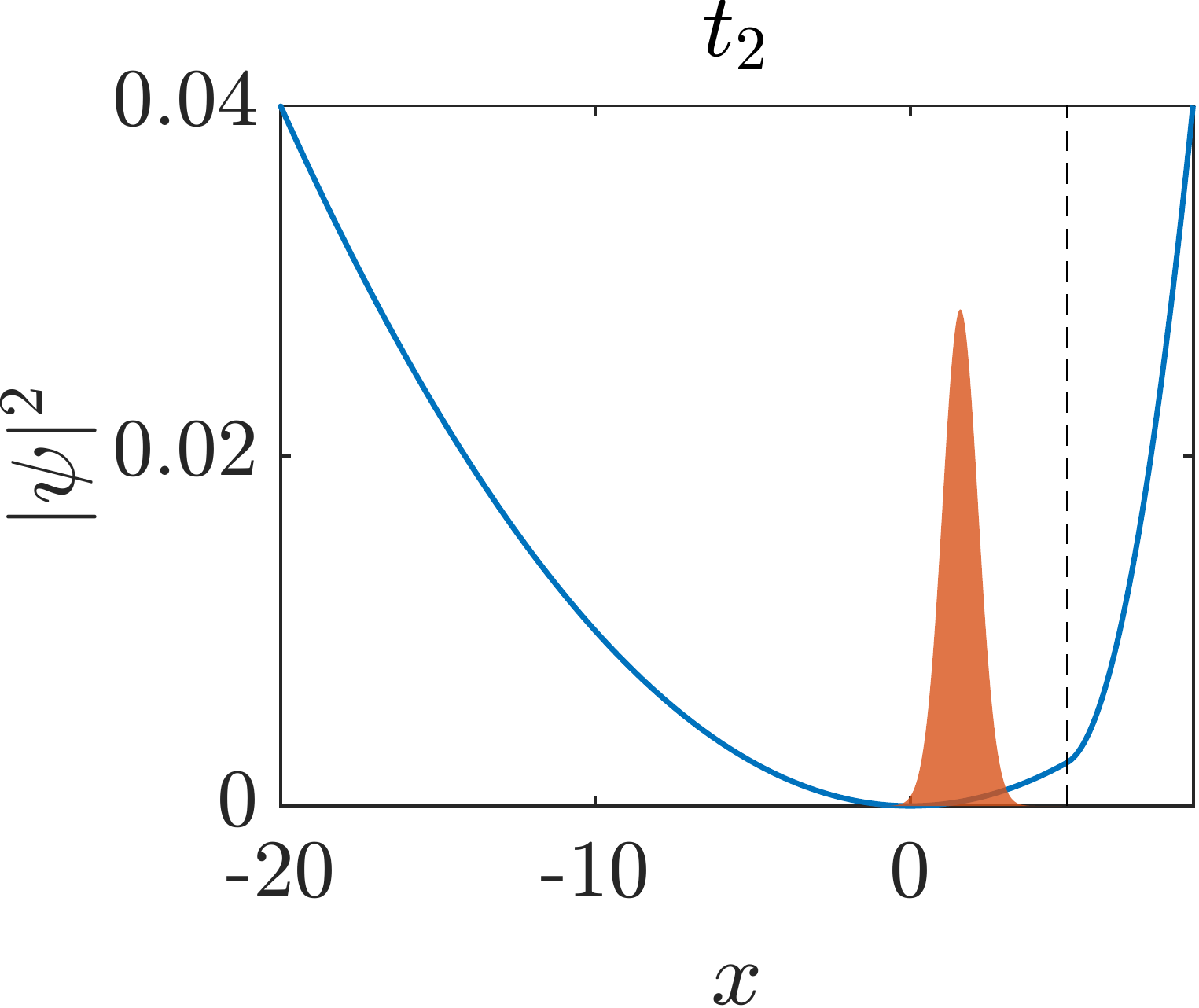}\hspace{0.01\linewidth}
\includegraphics[width=0.235\linewidth]{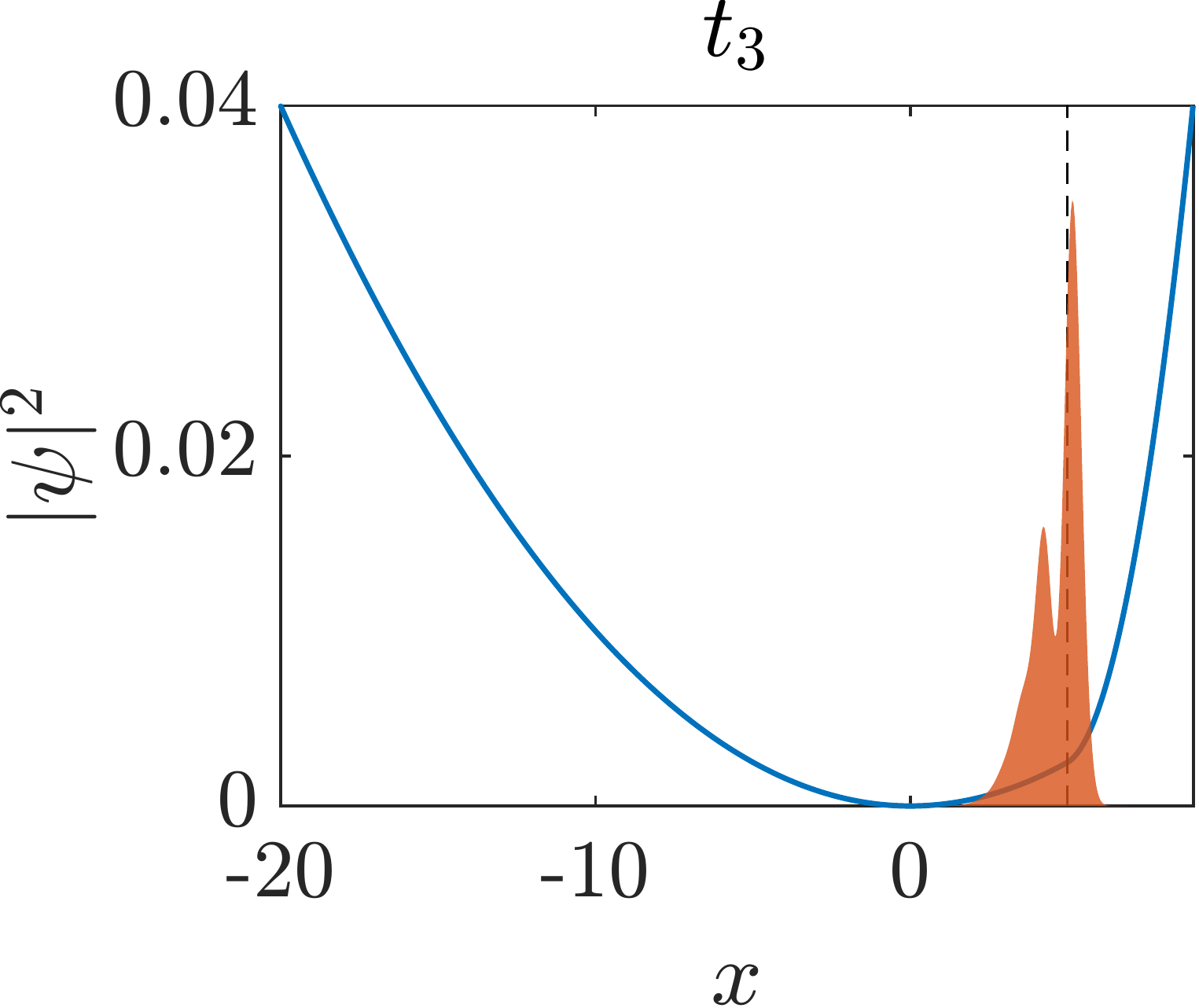}\hspace{0.01\linewidth}
\includegraphics[width=0.235\linewidth]{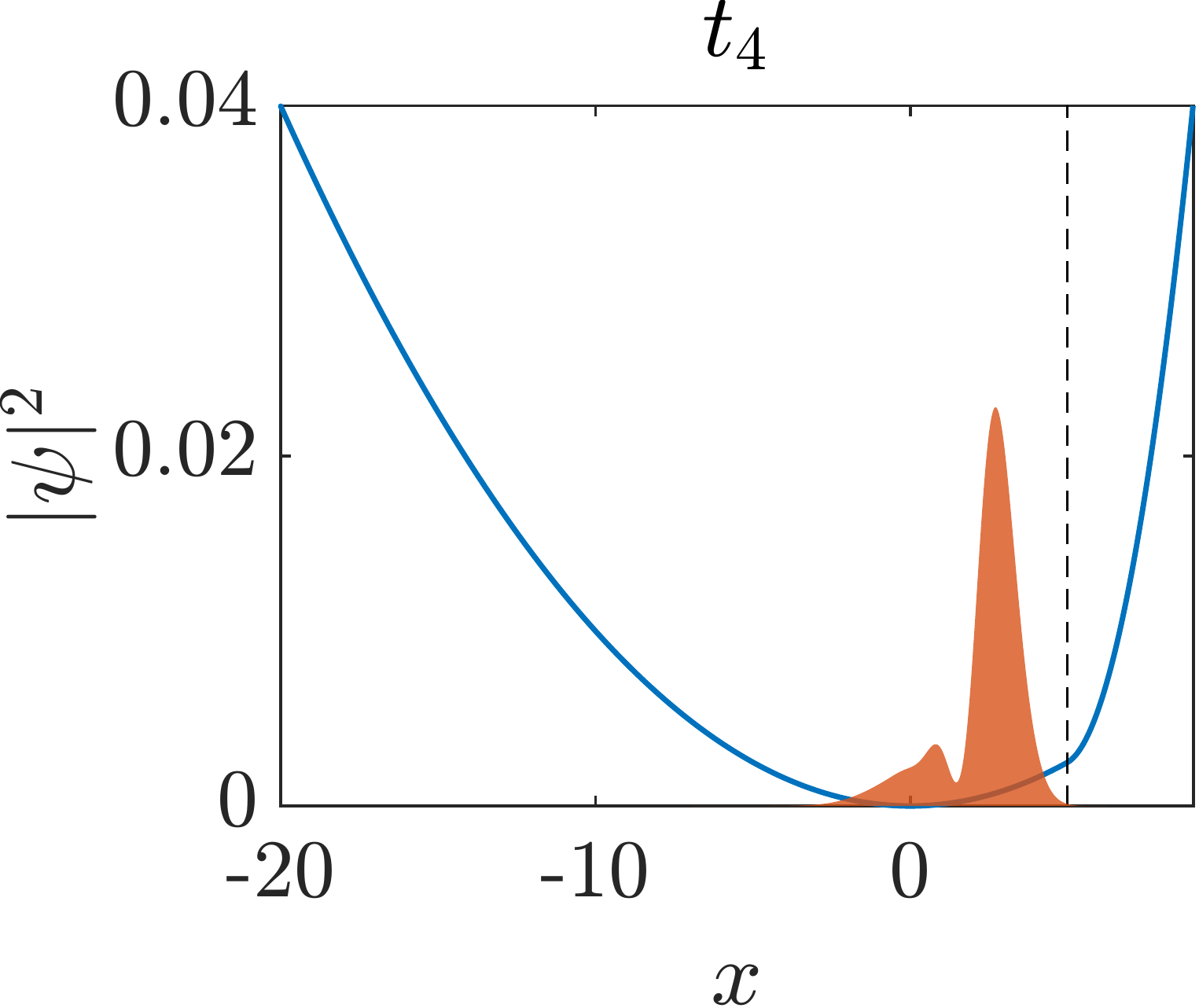}
\caption{Evolution of the probability density distribution of the
  quantum particle at four time instants, with
  $t_{1}<t_{2}<t_{3}<t_{4}$. Dashed line indicates the equilibrium
  position of the wall. The potential function, plotted in blue for
  the sake of visualization, is not to scale.}
\label{timeplots}
\end{figure*}

The quantum version of the above system will be a particle in a
potential well, which is the same as the harmonic oscillator potential
for $x \leq x_{\rm wall}$ and is given by a different parabolic
function for $x \geq x_{\rm wall}$. The potential function of
the system is given by 
\begin{equation}
    V(x) =
\left\{
	\begin{array}{ll}
		\frac{1}{2}k_{1}x^{2}  ,& x \leq x_{\rm wall} \\
		\frac{1}{2}k_{1}x^{2} + \frac{1}{2}k_{2}(x-x_{\rm wall})^{2} ,& x \geq  x_{\rm wall}
	\end{array}
\right.\label{softeq}
\end{equation}
We numerically solve the time-dependent Schr\"odinger equation for this system using the finite difference method by dividing the range $[-30,30]$ of the one-dimensional configuration space into 1,500 segments. We start from an initial wavefunction, which is a Gaussian function centered at $x=-5.0$, and standard deviation 1.0. This initial state corresponds, in the classical picture, to releasing the mass from the point $x=-5.0$, which would subsequently graze the wall located at $x=5.0$. The other parameters are taken as $m=1$, $k_1=1$, $k_2=10$. All quantities in this work are in units where $\hbar=1$. Fig.~\ref{timeplots} shows snapshots of the dynamics of the wavefunction for this system.

\begin{figure}
    \centering
    \includegraphics[width=\linewidth]{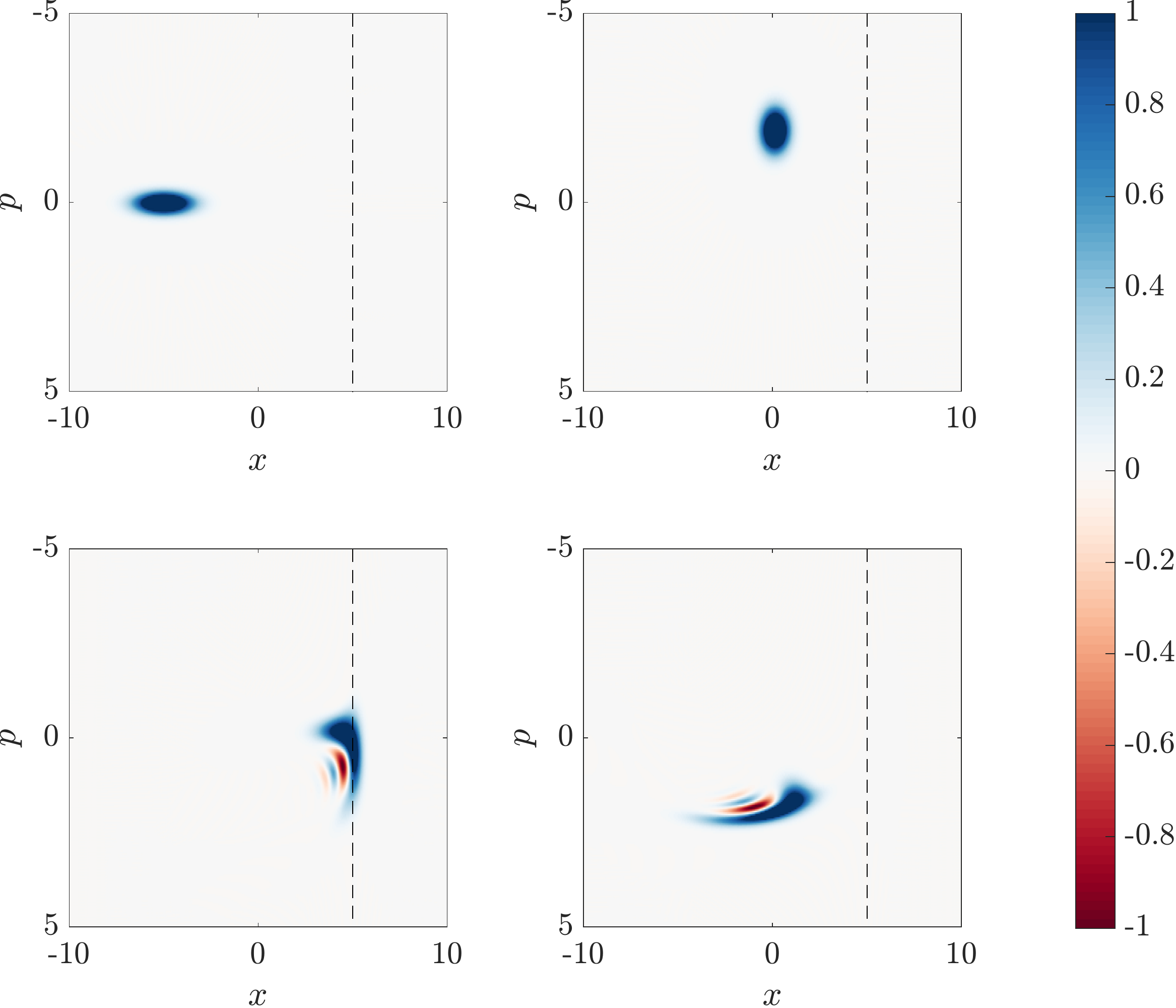}
    \caption{The evolution of the Wigner function at four time instants. The dashed line indicates the equilibrium position of the wall}
    \label{wigner}
\end{figure}

 To investigate the dynamics in the phase space, we compute the Wigner function
    \begin{equation}\label{eq:10}
        W(x,p)=\frac{1}{{\pi}\hbar}\int_{-\infty}^{\infty}\psi^{*}(x+y)\psi(x-y)e^{\frac{2ipy}{\hbar}}  \,dy
    \end{equation}
    which gives a real valued function of the position and momentum,
    which varies with time. Fig.~\ref{wigner} shows the plots of the
    Wigner function in the $x$--$p$ phase space at four different time
    instants. Prolonged observation of both $|\Psi^2|$ and the Wigner
    function shows that the time-evolution of the system is aperiodic.

    Although all information about the system is contained in the wavefunction, or equivalently the Wigner quasiprobability distribution, it is hard to characterize the type of dynamics using them. A time series of real values is much more amenable to such analysis. Other investigators have used the expectation value of an observable for this purpose \cite{ mendes1991sensitive, haake1992lyapunov, laha2020bifurcations}. Since different states can have the same expectation value for an observable, we have used a different quantity, the absolute value of the overlap  of the wavefunction at time $t$ with the initial wavefunction, given as $$\mathcal{O}(t)=\langle \psi(0)\vert \psi(t)\rangle .$$
    \begin{figure} [tbh]
        \centering
        \includegraphics[width=\linewidth]{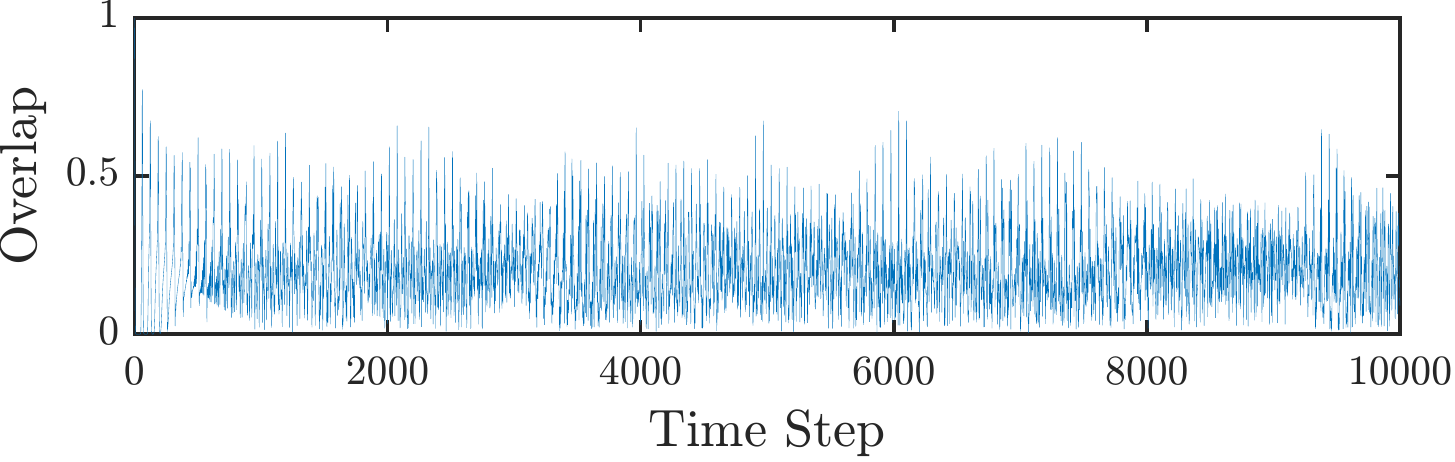}
        \caption{Plot of the overlap time series.}
        \label{timeseries}
    \end{figure}
        
\begin{figure}
        \centering
        \includegraphics[width=\linewidth]{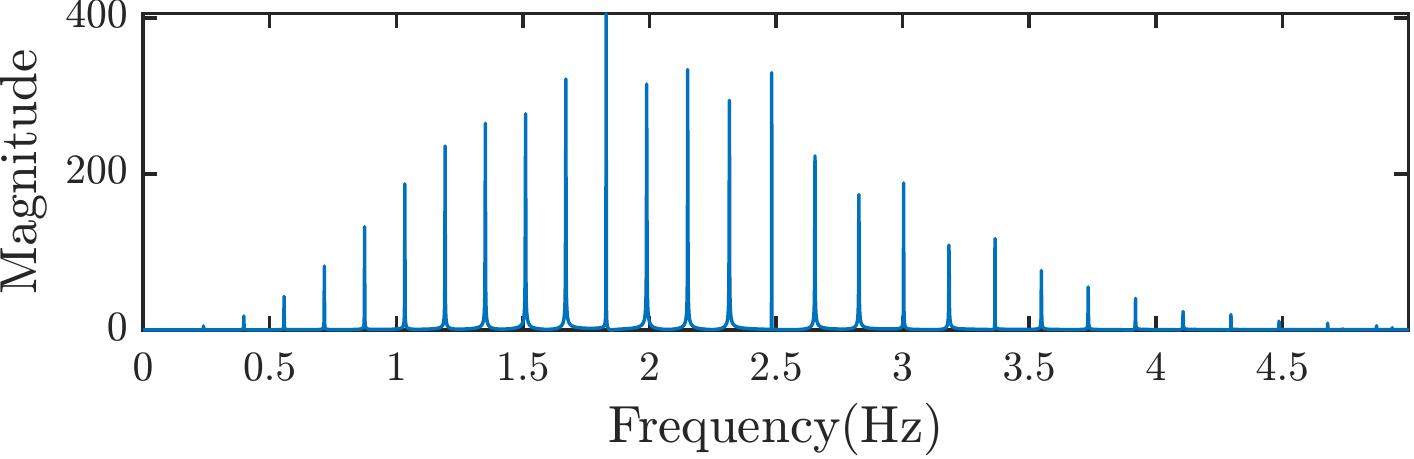}
        \caption{Frequency spectrum of the overlap time series.}
        \label{fft}
\end{figure}

We find that, when the wall is placed far away from the particle, the
behavior is periodic, like that of a harmonic oscillator. But when it
comes closer (even when the classical system would not make any
impact with the wall) the evolution of the wavefunction becomes
aperiodic. Fig.~\ref{timeseries} shows the time series of
$\mathcal{O}(t)$ for a wall position corresponding to the classical
grazing condition. We have performed the 0--1 test for chaos
~\cite{gottwald2009implementation}, and have found that the behavior
is not chaotic. A Fourier transform (Fig.~\ref{fft}) reveals that
there are many discrete frequencies in its dynamics.

Therefore, we conclude that, if the wavefunction is allowed to evolve
indefinitely following the Schr\"odinger equation, the evolution of
the wavefunction would be aperiodic, but is not chaotic as there is no
sensitive dependence on initial condition. Moreover, the frequency spectrum does
not have a continuum of frequencies and is a combination of a countable 
infinity of discrete frequencies.

\begin{figure} [tbh]
  \centering
\includegraphics[width=0.8\linewidth]{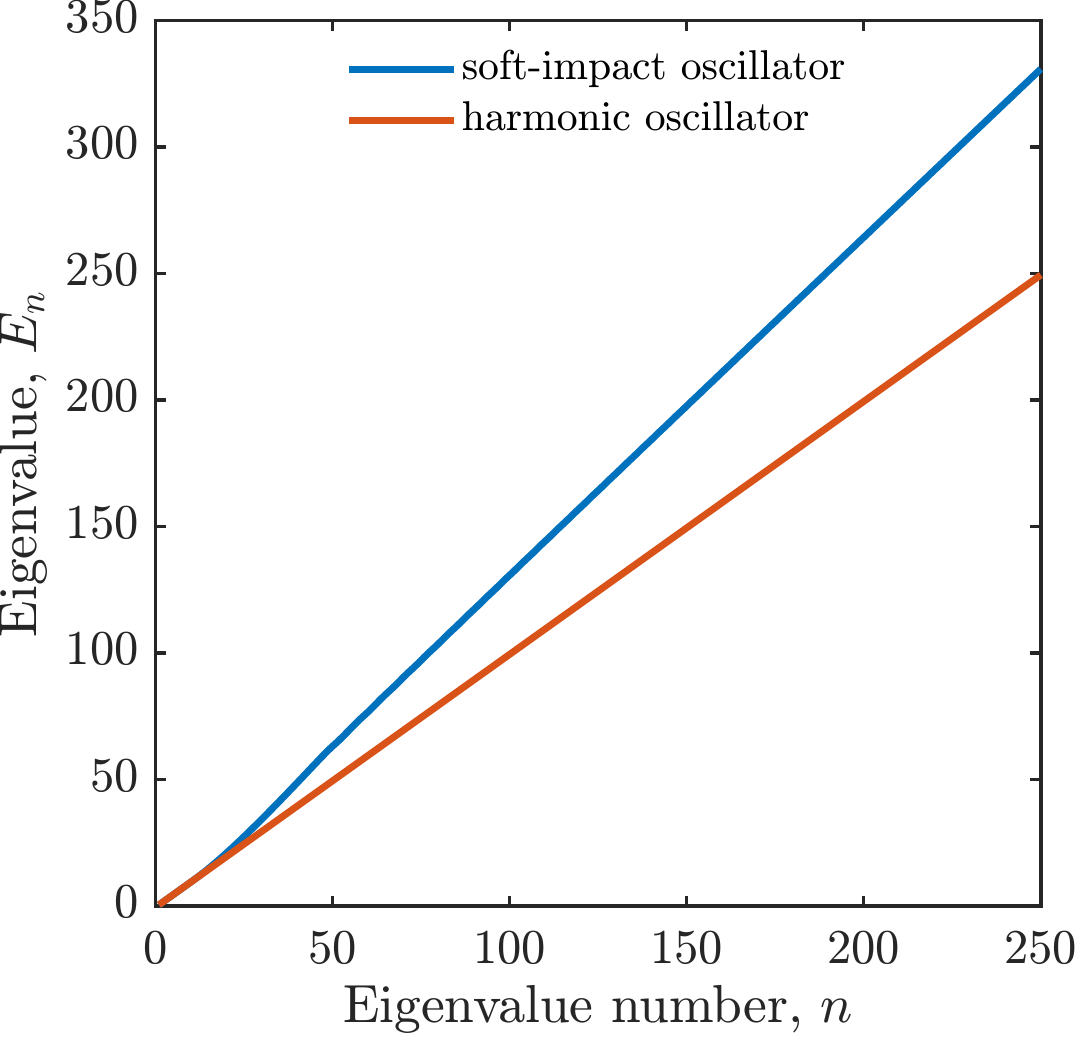}
\caption{A plot of the first 250 eigenvalues.}
\label{eigens}
\end{figure}

A plot of the eigenvalues (Fig.~\ref{eigens}) shows that for low
values of $n$, the eigenvalues coincide with those of the harmonic
oscillator. But for larger values of $n$, the graph has a different
slope because eigenfunctions corresponding to large
eigenvalues, which would usually be spread over a large region, cannot
extend far beyond the position of the wall. An energy
measurement can yield any of these eigenvalues and the expectation
value is computed to be 13.125~GeV for our choice of initial state.

\section{Collapse!}

In the last section we considered a situation where the wavefunction
evolves solely following the Schr\"odinger equation. In this section
we bring in a new possibility. Since the wall can be considered to be
a macroscopic object, an interaction of the particle with the wall
(classically, an impact) may amount to a position measurement, which
will cause the wavefunction to collapse. Following a collapse, the
wavefunction will continue to evolve according to the Schr\"odinger
equation until the next collapse. Thus, if this possibility is
considered, the evolution would contain unitary evolution as well as
non-unitary collapse processes.

Unlike the classical impact oscillator, the impact of the particle
with the wall will be a probabilistic event, guided by the
pre-collapse wavefunction of the particle. However, the present
knowledge does not allow us to pinpoint a unique algorithm with which
the instant of collapse and the location of the collapsed wavefunction
can be simulated. So we posit different postulates regarding the
mechanism of collapse, and work out the implication of each.

\begin{description}

\item[Postulate 1:] If the probability of finding the particle beyond the position of the wall exceeds a fraction $r$, i.e., if
  \begin{equation}
\int_{x_{\rm wall}}^\infty |\Psi|^2 dx \geq r,\;\;\;\;\;\;r \in (0,1)
  \end{equation}
  then the wavefunction collapses to the position of the wall.

\item [Postulate 2:] The same as Postulate 1, except that the number $r$ is not fixed, and is a random number between 0 and 1.

\item [Postulate 3:] The same as Postulate 1, except that the
  wavefunction collapses to a point given by the pre-collapse
  probability distribution.

\item [Postulate 4:] The same as Postulate 2, except that the
  wavefunction collapses to a point given by the pre-collapse
  probability distribution.

\end{description}

\begin{figure} 
  \centering
\includegraphics[width=0.48\linewidth]{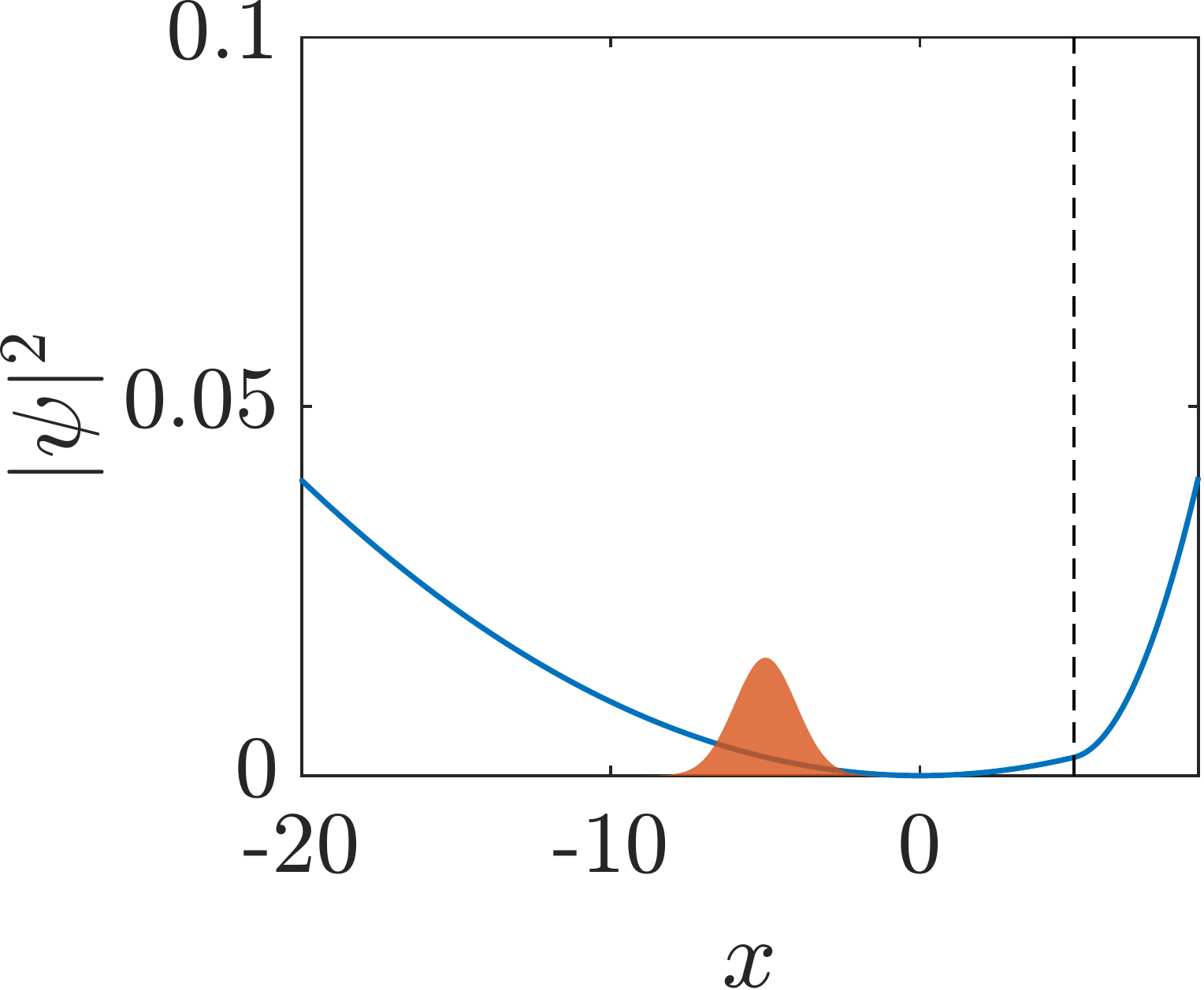}
\hspace{0.001\linewidth}
\includegraphics[width=0.48\linewidth]{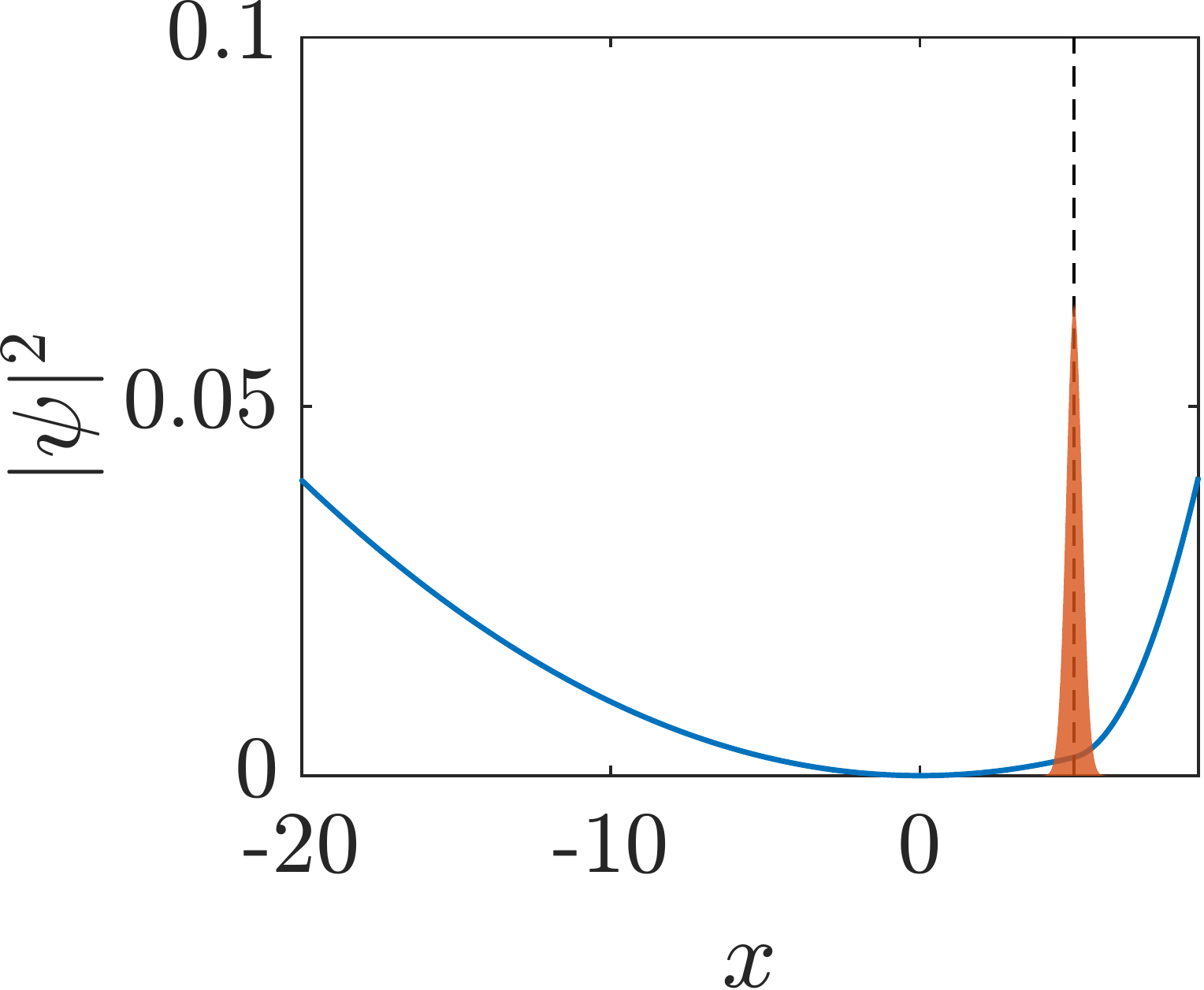} \\
\hspace{0.1\linewidth}{\footnotesize (a)}  \hspace{0.48\linewidth}{\footnotesize (b)}
\caption{(a) Initial Gaussian with mean $-5.0$ and SD=1, (b) the post-collapse wavefunction---a narrow Gaussian located at the position of the wall $x=5.0$ and SD=0.25. The potential function is shown in blue (not to scale).}
\label{collapse}
\end{figure}

In the following simulations, the initial wavefunction is considered
to be a Gaussian function centred at $x=-5.0$ with standard deviation
1. The wall is placed at $x=5.0$, i.e., where the classical oscillator
would experience grazing.  For the first and third postulates, the value of $r$ is taken as 0.5. The post-collapse wavefunction is supposed
to be an eigenfunction of the position operator, i.e., a delta
function. However, the numerical routine would not work with such
discontinuous functions. So we consider the post-collapse wavefunction
to be a narrow Gaussian function of standard deviation 0.25
(Fig.~\ref{collapse}). The parameter values are taken as: mass of the
quantum particle $m = 1$, spring constant of the spring attached
to the mass $k_{1} = 1$, the spring constant corresponding
to the soft wall $k_{2}=10$, time step $\delta t=
0.1$. We calculate for a total number of 10,000 time
steps. The first 150 eigenfunctions have been used for the time
evolution.

\section{Results}

Our objective here is to obtain testable predictions of the various
possible mechanisms of evolution of the wavefunction as outlined
above. We focus on two observables: energy and position.

\subsection{Probability distribution of energy values}

\begin{figure} 
\centering
\includegraphics[width=0.85\linewidth]{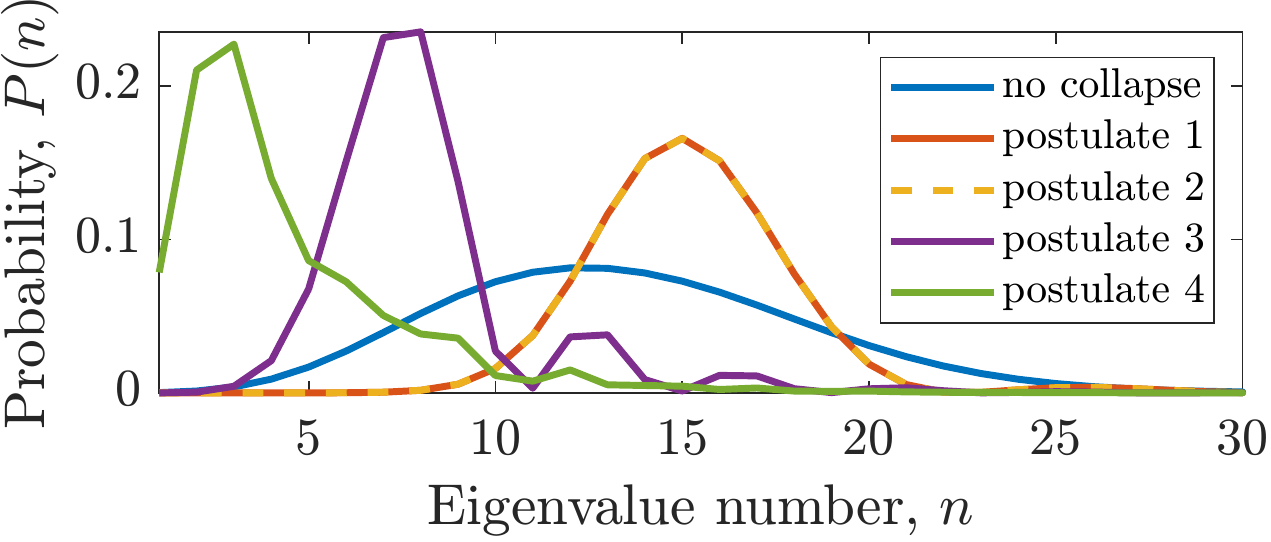}
\caption{Comparison of probability distributions for energy eigenvalues for the different collapse postulates.  }
\label{eigprobs}
\end{figure}

An energy measurement may return any energy eigenvalue presented in
Fig.~\ref{eigens}, but the probability of finding each eigenvalue would
be different for various postulated situations. For the different
postulates, the computed probability distribution of the energy eigenvalues are
presented in Fig.~\ref{eigprobs} and the expectation values of energy
are tabulated in Table~\ref{tab:expE}.

\begin{table}[htp]
    \centering \small
    \begin{tabular}{@{}llll@{}}
        \toprule
        \textbf{Postulate}       &&&\textbf{Expected energy}\\
        \hline 
        No collapse	    &&&13.125~GeV\\
        Postulate 1     &&&14.75~GeV\\
        Postulate 2     &&&14.62~GeV\\
        Postulate 3     &&&11.46~GeV\\
        Postulate 4	    &&&5.57~GeV\\
        \bottomrule
    \end{tabular}
    \caption{Expectation values of energy in the five postulated situations.}
    \label{tab:expE}
\end{table}

Note that the expectation value of energy (Table \ref{tab:expE}) for
unitary evolution of the wavefunction depends on the initial
wavefunction considered and those for the four collapse postulates
depend on the variance of the post-collapse wavefunction. Since we
have considered the standard deviation to be 0.25 in all cases, one
should pay attention to the relative magnitudes rather than the
absolute magnitudes of the expectation values.

The results presented in Fig.~\ref{eigprobs} and
  Table~\ref{tab:expE} show that, if experiments give the expectation
  value of energy larger than what is predicted by standard quantum
  mechanics for this system, then Postulate 1 or 2 may be true, and if
  they give a lower value, Postulate 3 or 4 may be true.

\subsection{Probability distribution of position values}

\begin{figure}[tbh]
  \centering
  \raisebox{0.6in} {\footnotesize (a)} \hfill \includegraphics[width=0.75\linewidth]{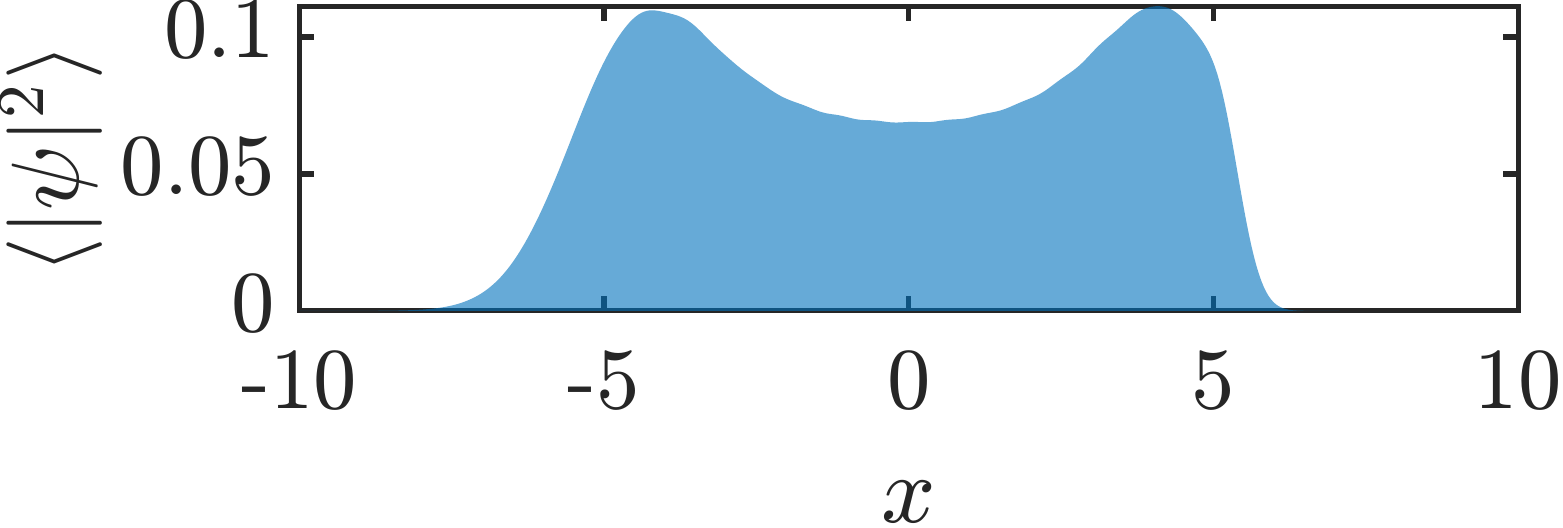} \\ \vspace{0.2in}
  \raisebox{0.6in} {\footnotesize (b)} \hfill \includegraphics[width=0.75\linewidth]{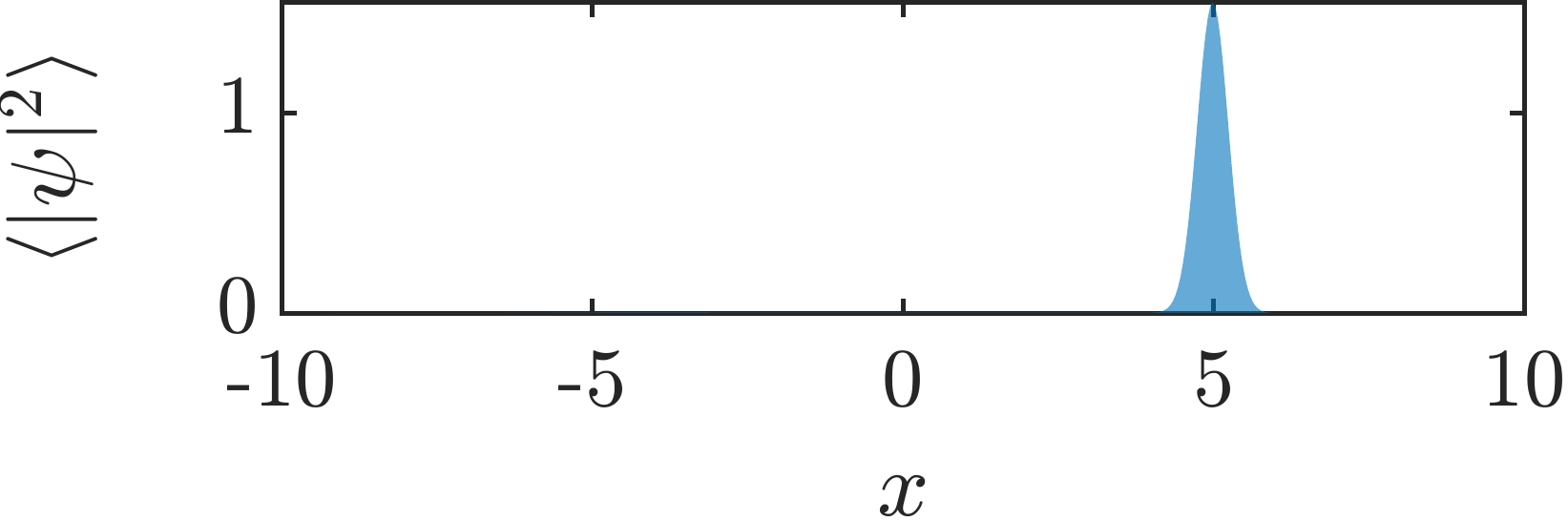}\\\vspace{0.2in}
  \raisebox{0.6in} {\footnotesize (c)} \hfill \includegraphics[width=0.75\linewidth]{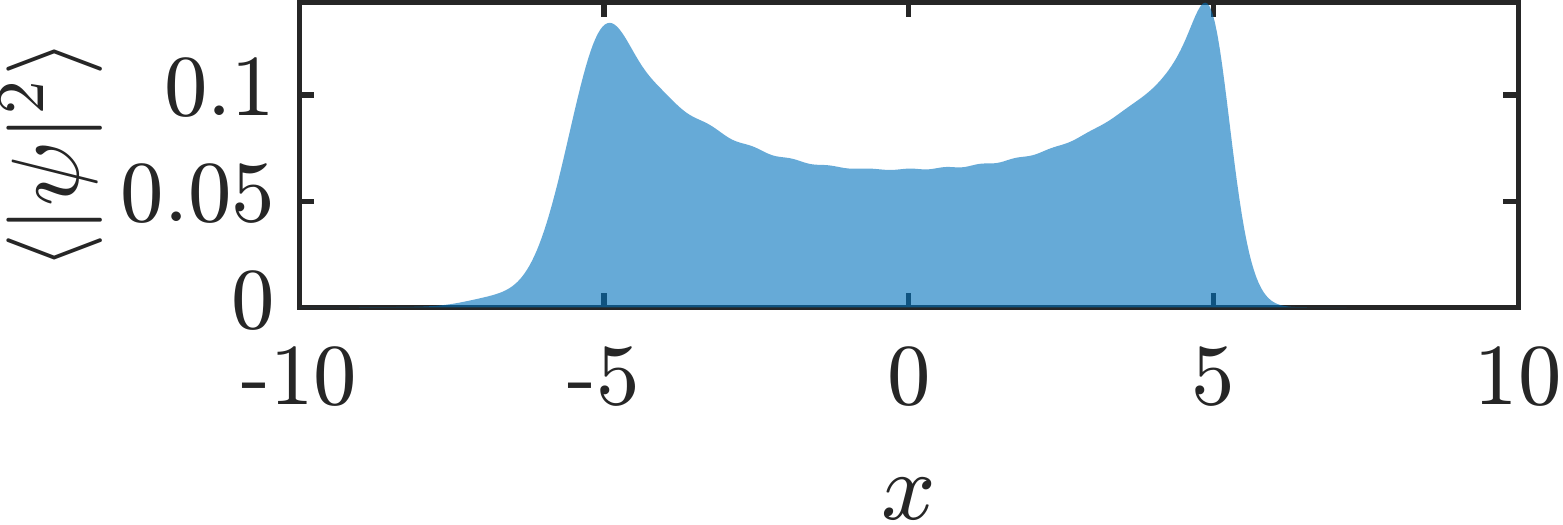} \\ \vspace{0.2in}
  \raisebox{0.6in} {\footnotesize (d)} \hfill \includegraphics[width=0.75\linewidth]{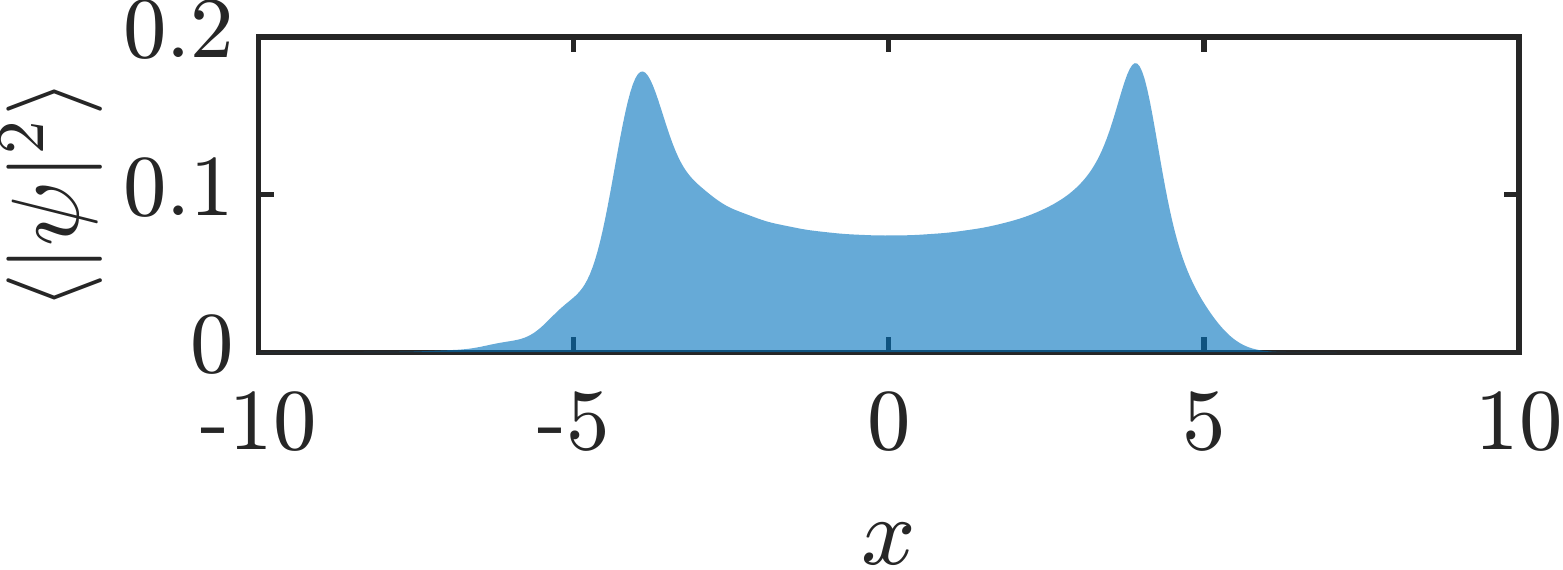}\\\vspace{0.2in} 
  \raisebox{0.6in} {\footnotesize (e)} \hfill \includegraphics[width=0.75\linewidth]{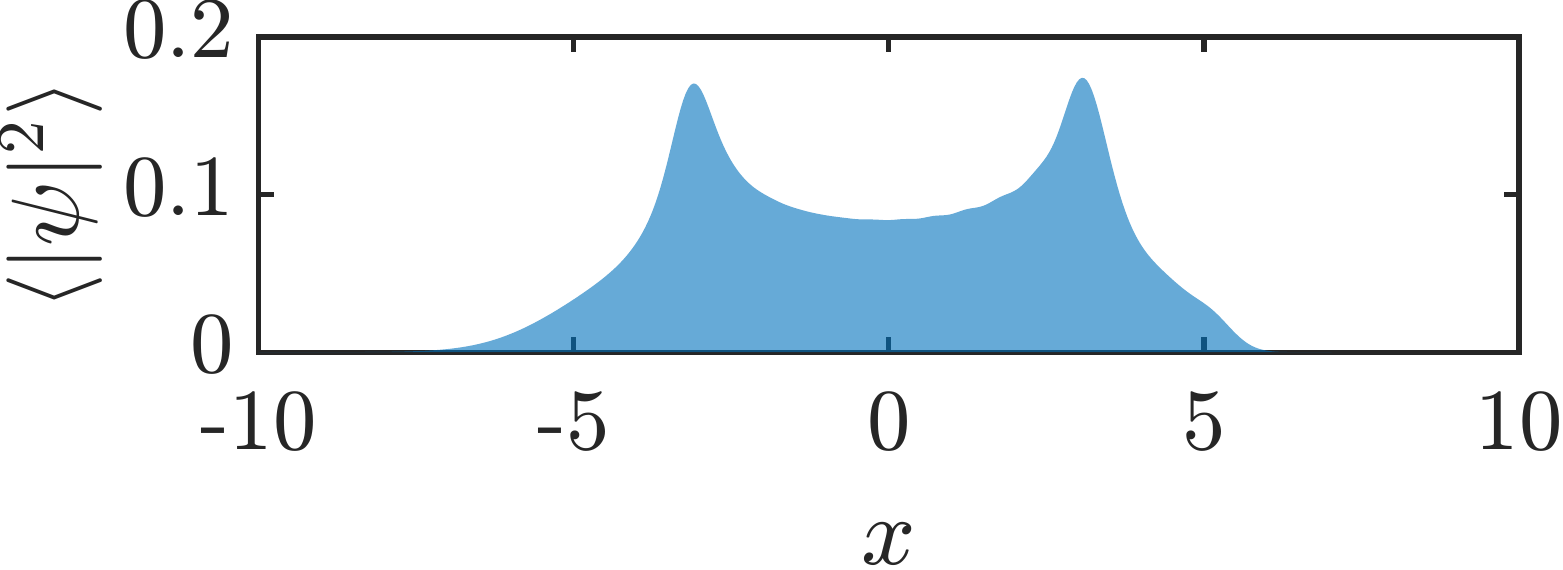}
\caption{The averaged probability density functions of the position of the particle
  (a) without collapse, (b) for postulate 1, (c) for postulate 2, (d) for postulate 3, and (e) for postulate 4.} 
\label{pos-dist}
\end{figure}

We have computed the distribution of position values obtained through
10,000 time-steps. These are plotted in Fig.~\ref{pos-dist}.

\begin{table} 
    \centering \small
    \begin{tabular}{@{}lllllll@{}}
        \toprule
        \textbf{Postulate}       &&&\textbf{Mean}      &&&\textbf{SD}\\
        \hline 
        No collapse	    &&&-0.2662	&&&3.4963\\
        Postulate 1     &&&4.8217	&&&1.1296\\
        Postulate 2     &&&-0.2630	&&&3.6354\\
        Postulate 3     &&&-0.0657	&&&3.1330\\
        Postulate 4	    &&&-0.0812	&&&2.8818\\
        \bottomrule
    \end{tabular}
    \caption{Mean and standard deviations of the time averaged PDFs for the different collapse postulates.}
    \label{tab:pdf_position}
\end{table}

It is found that if Postulate 1 is true, there is only one peak in the probability distribution and in all other cases there are two peaks. For unitary evolution without collapse, the two peaks are of almost the same height while in the other cases they are of dissimilar heights.
Table~\ref{tab:pdf_position} gives the mean and standard deviation of the distributions for the five cases shown in Fig.~\ref{pos-dist}.

\section{The wall represented by another particle}

In the previous section, we considered how a macroscopic object like a
spring-supported soft wall might bring about the collapse of the
wavefunction of a particle. We now ask: What happens if the wall is
represented by another microscopic particle?
Can such an interaction between two quantum particles lead to collapse
of their wavefunctions?

\begin{figure} 
  \centering
  \includegraphics[trim={7cm 5.5cm 7cm 8.35cm},clip,width=0.7\linewidth]{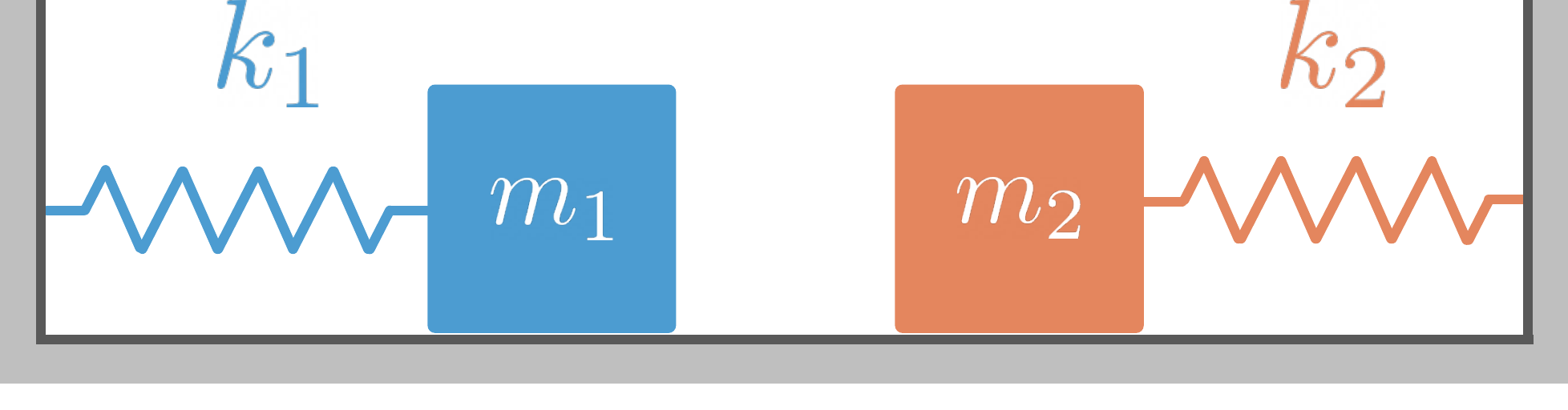}
  \caption{Classical analog of the two-particle system. \label{two-dist}}
\end{figure}

We consider the soft wall to be made up of a particle of mass $m_2$,
oscillating in its own harmonic potential with spring constant $k_2$
(Fig. \ref{two-dist}). Classically, this corresponds to the situation
where the soft wall is no longer massless. Each particle is in their
own harmonic potential wells and there is no interaction between them
except when they meet. We assume that they may meet when their
wavefunctions have a significant overlap. We postulate that an
interaction of this kind can also lead to the collapse of their
wavefunctions, and work out the implications of such an assumption.

 The Hamiltonian of the system is 
\begin{equation}
      \hat{H}= -\frac{\hbar^{2}}{2m_{1}}\frac{\partial^{2}}{\partial{x_{1}}^{2}} - \frac{\hbar^{2}}{2m_{2}}\frac{\partial^{2}}{\partial{x_{2}}^{2}} + \frac{k_{1}x_{1}^{2}}{2} + \frac{k_{2}x_{2}^{2}}{2}
\end{equation}
where $k_1$ and $k_2$ are the spring constants of the two harmonic oscillator potentials; $m_1$ and $m_2$ are the masses of the two particles. Since there is no interaction term in this Hamiltonian, the state of the two-particle system remains separable: 
$$\psi(x_1,x_2)=\psi_1(x_1)~\psi_2(x_2)$$

The evolution is governed by their single-particle Schr\"odinger equations
\begin{equation}
    i \hbar \frac{\partial \psi_1}{\partial t} = \left[ -\frac{\hbar^{2}}{2m_{1}} \frac{\partial^{2}}{\partial{x_{1}}^{2}} + \frac{k_{1}x_{1}^{2}}{2}\right]\psi_{1}  
\end{equation}
and
\begin{equation}
    i \hbar \frac{\partial \psi_2}{\partial t} = \left[ -\frac{\hbar^{2}}{2m_{2}} \frac{\partial^{2}}{\partial{x_{2}}^{2}} + \frac{k_{2}x_{2}^{2}}{2}\right]\psi_{2}  
\end{equation}
except at instants when the wavefunctions collapse.

For numerical simulations, the harmonic potentials are taken to be
centered at $-2$ and $+2$, and the initial states of the two particles are
Gaussians centred at $-5$ and $+5$ respectively with a standard
deviation of 1.

\subsection{No collapse}
\begin{figure}[tbh]
    \centering
    \includegraphics[width=0.48\linewidth]{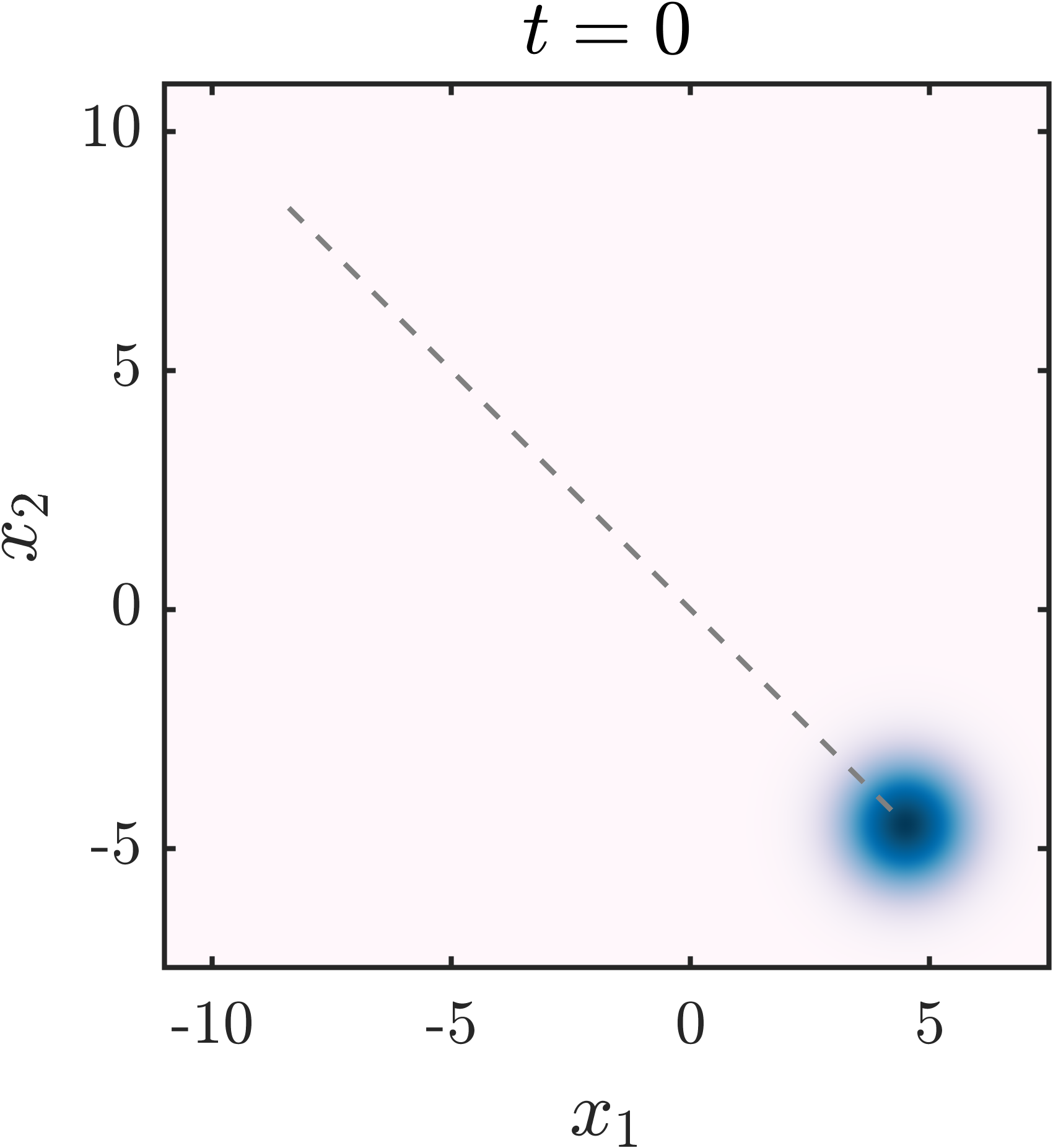} \hspace{0.005\linewidth} \includegraphics[width=0.48\linewidth]{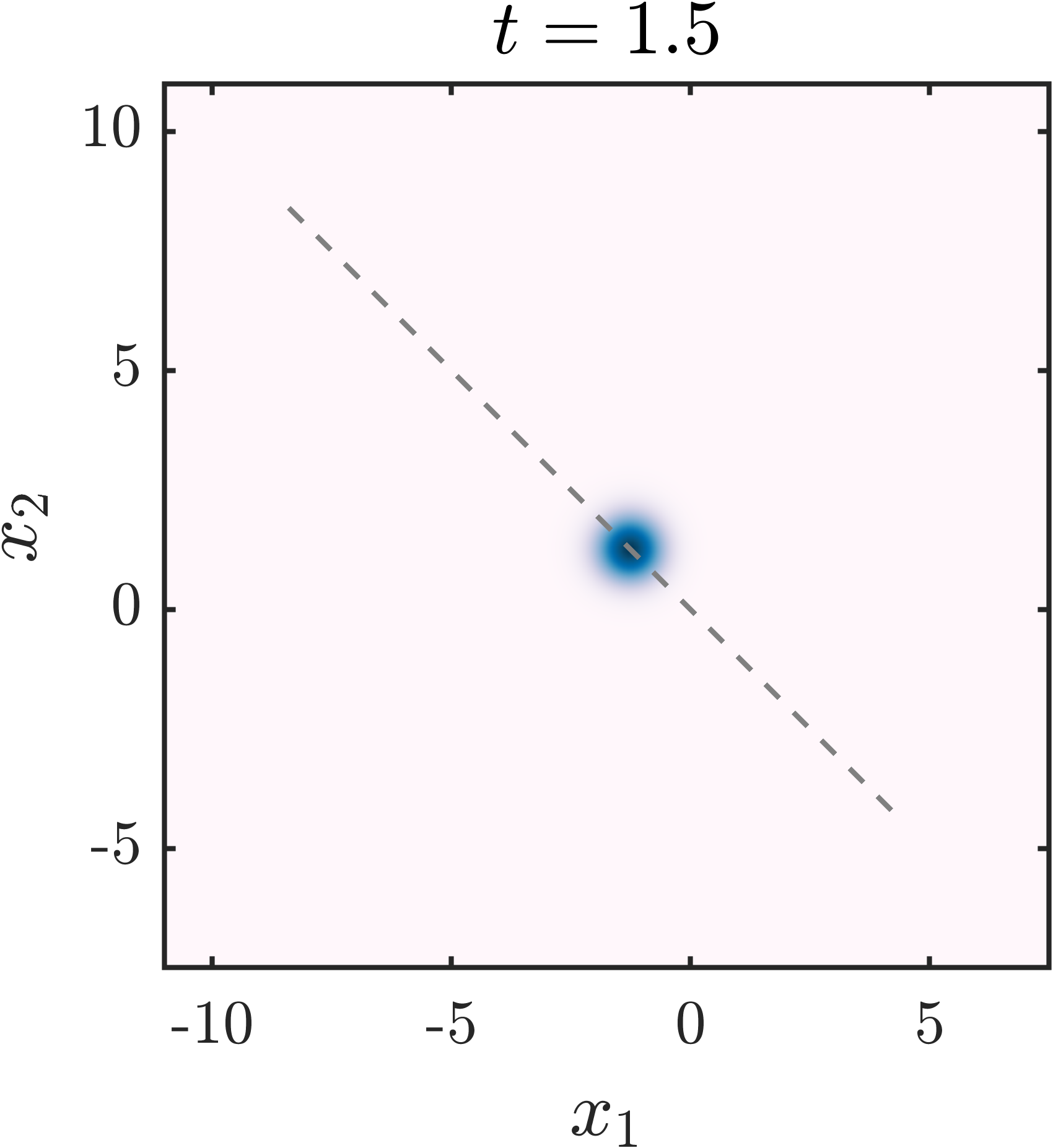} \\ \vspace{0.005\linewidth} \includegraphics[width=0.48\linewidth]{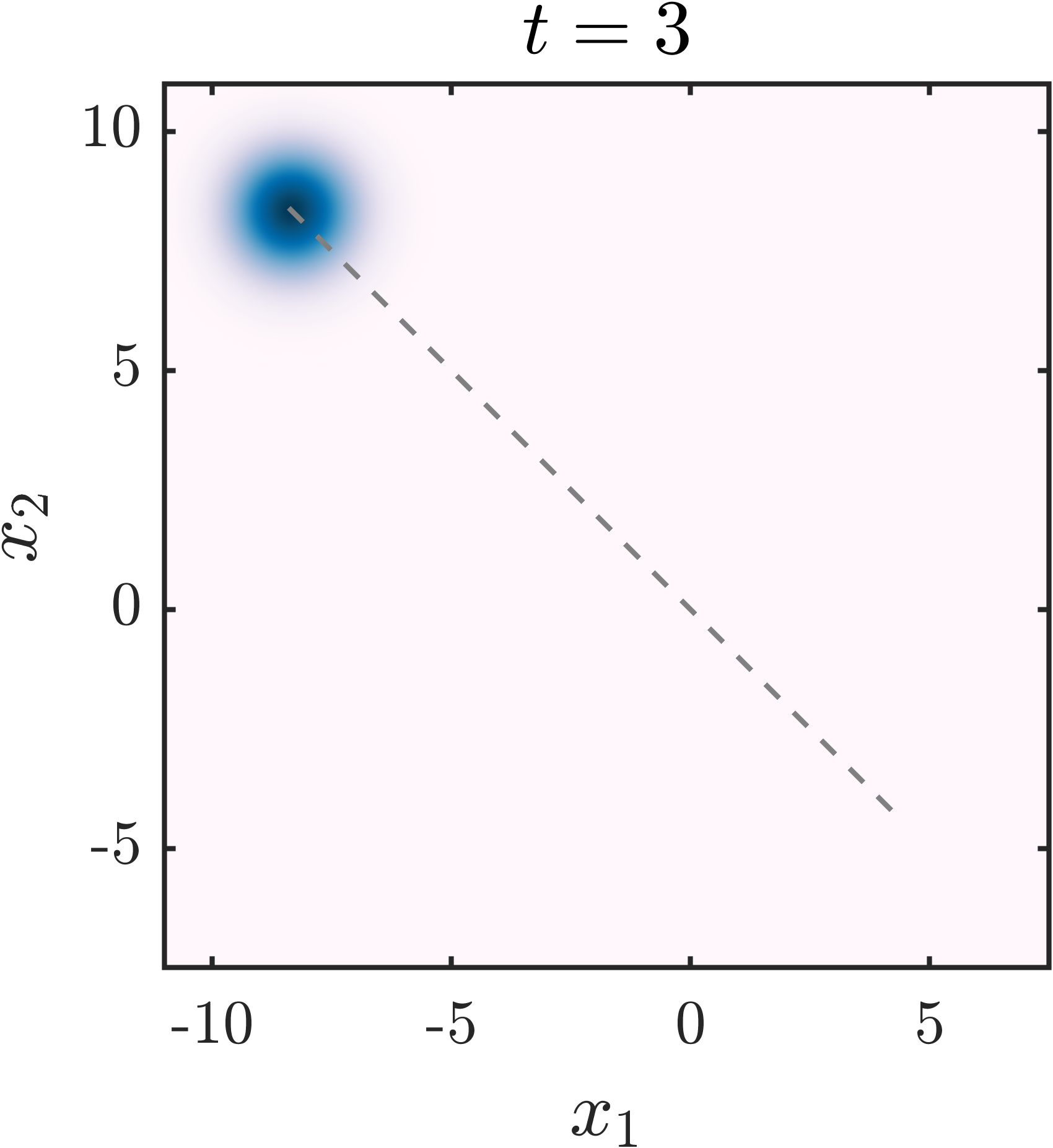} \hspace{0.005\linewidth} \includegraphics[width=0.48\linewidth]{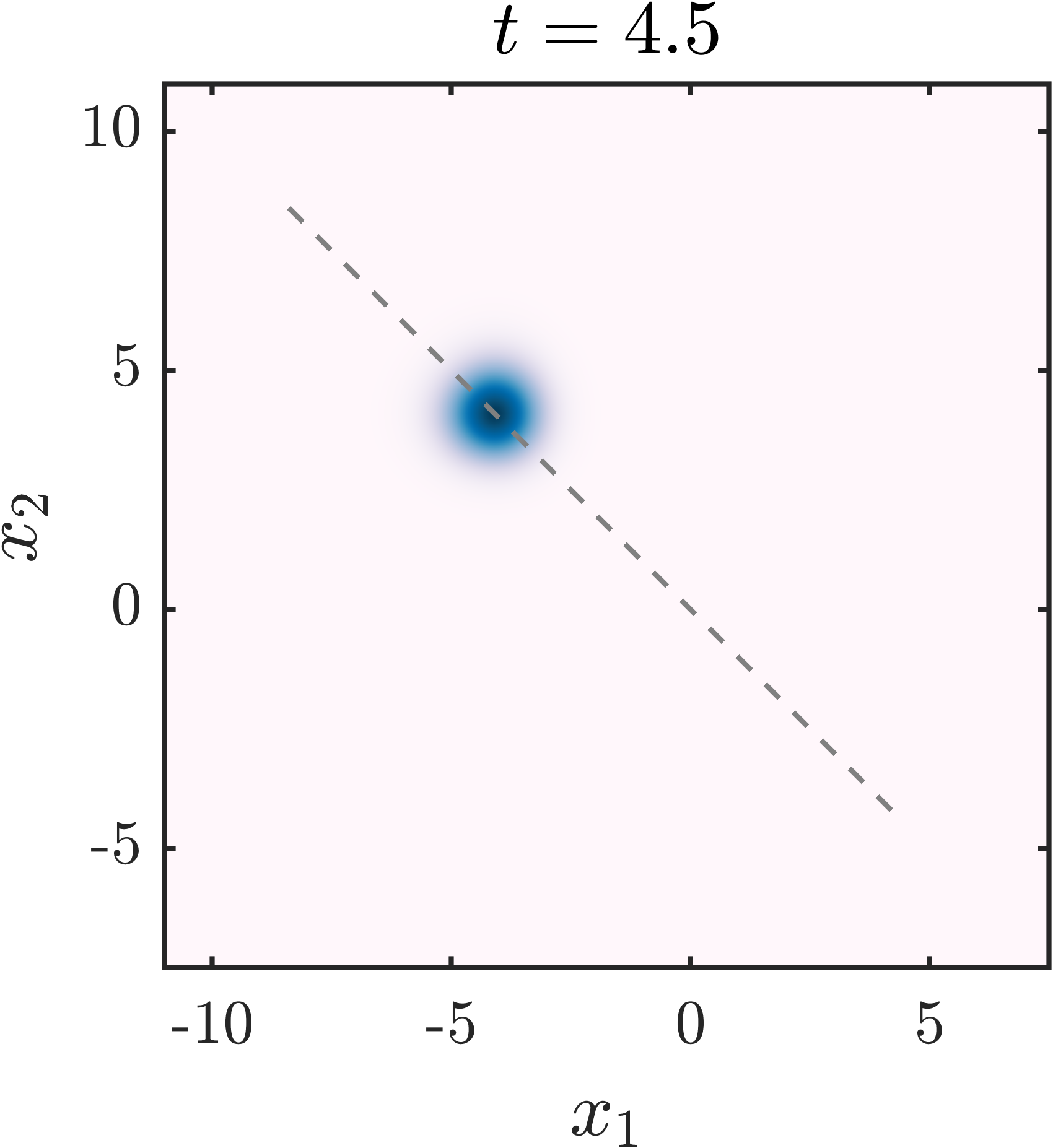} \caption{Evolution of the two-particle probability density $\vert\psi(x_1,x_2)\vert^2$ at four time instants for the no-collapse case. The dashed line is the locus of the expected position. }
    \label{contour}
\end{figure}

According to standard quantum mechanics, a system collapses into an
eigenstate only when a measurement takes place. So, in the scenario we
contemplate, the standard formulation predicts no collapse. To compare
our collapse theories with the orthodox quantum mechanical prediction,
we first simulate the system without wavefunction collapse. The
contour plots of the probability distribution ($|\psi(x_1,x_2)|^2$) for
such an evolution can be seen in Fig.~\ref{contour}.  The
distribution periodically spreads and shrinks while its mean value
moves along the dashed line shown. The evolution is completely
periodic for rational ${\omega_1}/{\omega_2}$ (where $\omega =
\sqrt{{k}/{m}}$) and quasiperiodic otherwise.

\subsection{Energy conserved collapse}
A shortcoming of orthodox quantum mechanics and dynamical collapse
theories like GRW and CSL is that they violate conservation of
energy. Dynamical collapse theories make small but testable
predictions for these violations \cite{goldwater2016testing} which can
be used to validate and narrow down the free parameters in these
models. Since no such violation has yet been reported, we invoke
Occam's razor and assume that energy is indeed conserved. We hence postulate two different collapse processes that
respect the conservation of energy. The first of the two processes
conserves energy of each individual particle, while the second
conserves the total energy of the system through the collapse process. These
processes are detailed below.
The expectation value of energy is
given by
$$\langle E \rangle = \langle \psi \vert  \hat H  \vert \psi \rangle = \int_{-\infty}^{\infty}  \psi^*(x)\, \hat H\, \psi(x)\, \mathrm{d}x $$
For a particle described by a Gaussian wavepacket centered at $a$ with SD 
\begin{figure}
    \centering
    \includegraphics[width=0.6\linewidth]{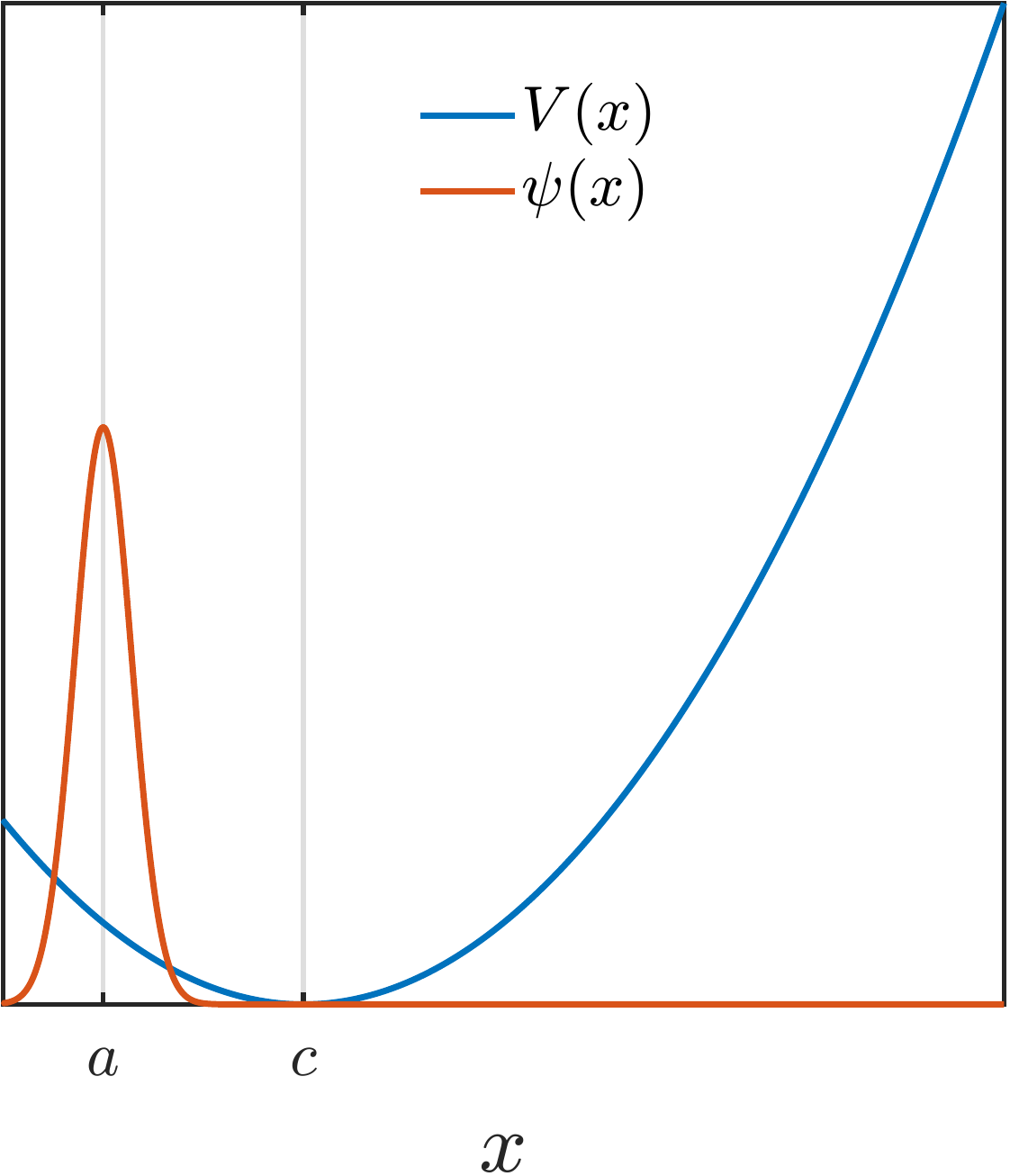}
    \caption{A Gaussian wavepacket in a translated harmonic potential. The centers of the wavepacket and the potential are $a$  and $c$ repectively.}
    \label{fig:init_potential}
\end{figure}
$\sigma$
$$\psi (x)=\frac{1}{\sqrt{\sqrt{2 \pi } \sigma }} \exp \left[{-\frac{(x-a)^2}{4 \sigma ^2}}\right]$$
in the translated harmonic potential (Fig.~\ref{fig:init_potential}) for which the Hamiltonian is
$$\hat H=-\frac{\hbar^2}{2m}\frac{\partial^2 }{\partial
  x^2}+\frac{1}{2}k(x-c)^2,$$ the expectation value of energy is
  \begin{widetext}
     \begin{align*}
         \langle E \rangle =& \int_{-\infty}^{\infty}  \psi^*(x)\, \hat H ~\psi(x)\, \mathrm{d}x\\  =& \int_{-\infty}^{\infty}  \frac{1}{\sigma \sqrt{2 \pi } } \mathrm{e}^ {-\frac{(x-a)^2}{4 \sigma ^2}}~ \left[-\frac{\hbar^2}{2m}\frac{\partial^2 }{\partial x^2}+\frac{1}{2}k(x-c)^2\right] ~ \mathrm{e}^ {-\frac{(x-a)^2}{4 \sigma ^2}}~ \mathrm{d}x\\
         =& \int_{-\infty}^{\infty}  \frac{1}{2\sigma \sqrt{2 \pi } } \mathrm{e}^ {-\frac{(x-a)^2}{2 \sigma ^2}}~ \left[-\frac{\hbar^2 \left\{ (x-a)^2 - 2\sigma^2\right\}}{4m\sigma^4}+ k(x-c)^2\right]~ \mathrm{d}x\\
     \end{align*}
  \end{widetext}
which can be integrated using Gaussian integrals to give
$$\langle E \rangle = \frac{k}{2} \left\{ \sigma^2 + (a-c)^2 + \frac{\hbar^2}{4mk\sigma^2} \right\}.$$

\begin{figure}[tbh]
    \centering
    \includegraphics[height=0.7\linewidth]{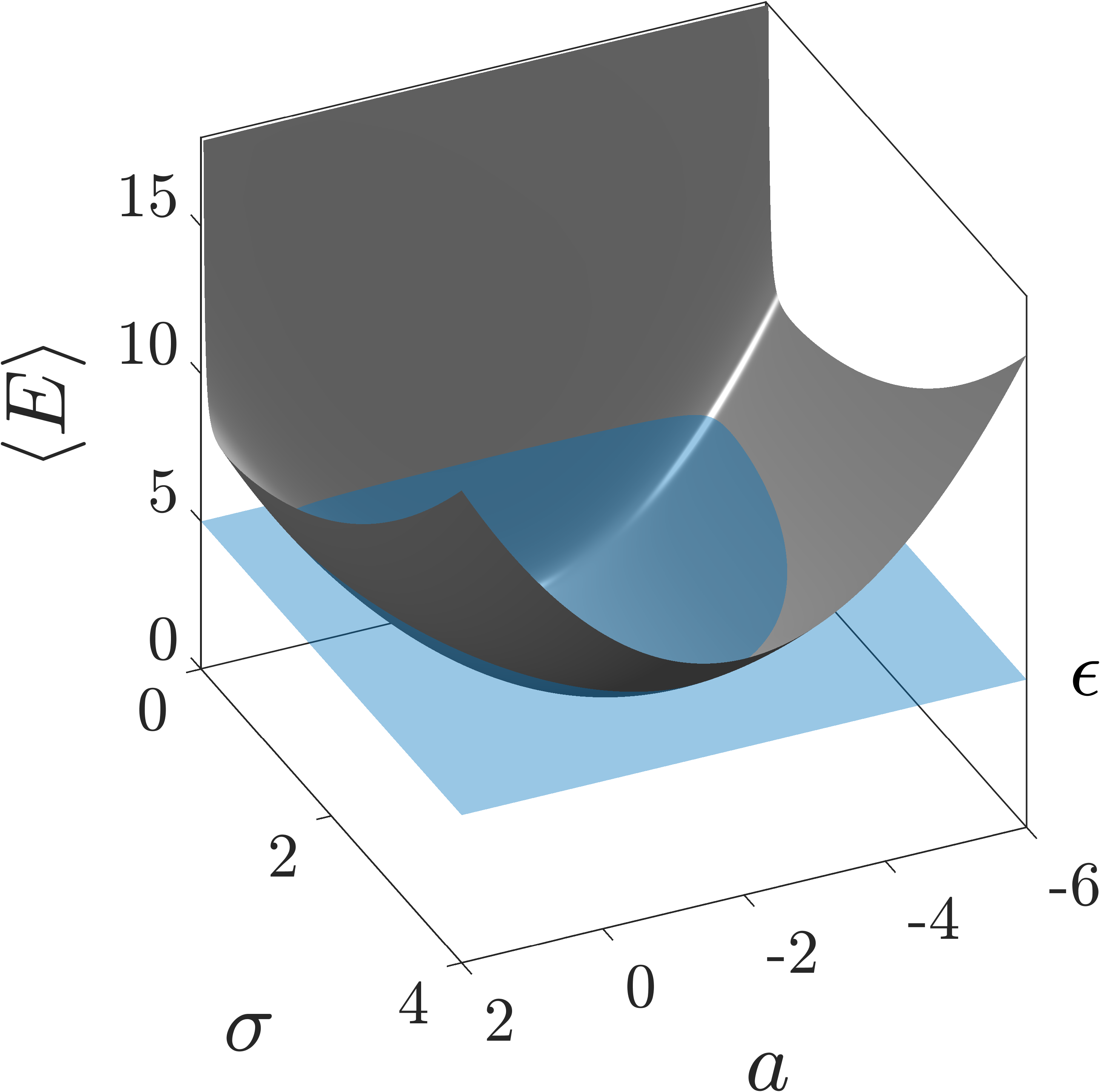} \\ \hspace{0.07\linewidth}(a) \\ \vspace{0.1\linewidth}
    \includegraphics[height=0.6\linewidth]{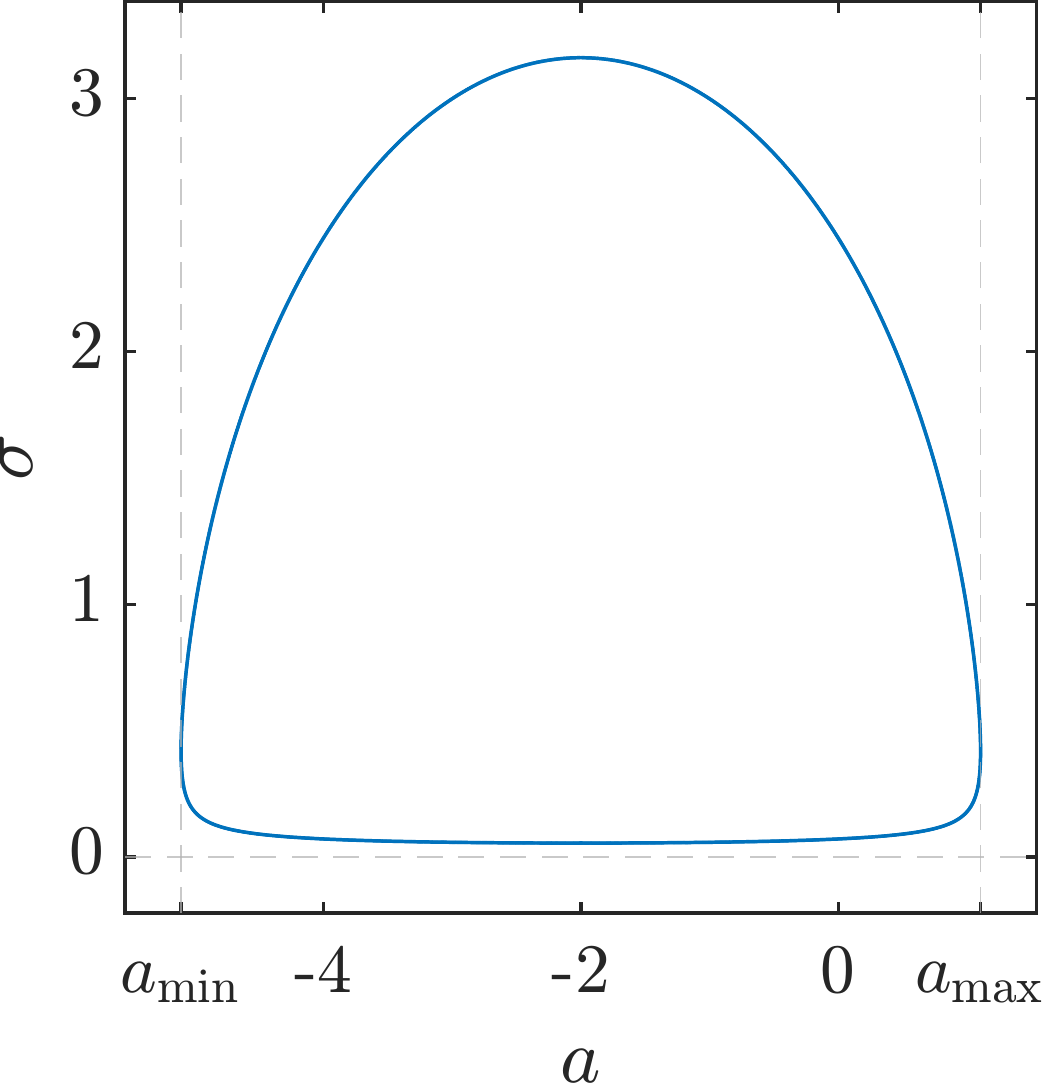}\\  \hspace{0.07\linewidth}(b)   
    \caption{Dependence of the expected energy $\langle E \rangle$ on the mean position $a$ and standard deviation $\sigma$ of the Gaussian wavepacket (gray) and surface of constant energy  $\epsilon$ (blue) (a), which intersects the former surface along the curve in (b). $k=m=1$, $c=-2$.}
    \label{fig:expectedEnergyconstraint}
\end{figure}

The dependence of the expected energy on the mean $a$ and the standard
deviation $\sigma$ of the Gaussian wavepacket is plotted in
Fig.~\ref{fig:expectedEnergyconstraint}(a). If the expected energy is
constrained to a certain value, say $\epsilon$, then the possible
values of $a$ and $\sigma$ get constrained to the intersection between
the surfaces $\langle E \rangle (a,\sigma)$ and $\langle E \rangle =
\epsilon$. The resulting family of curves parametrized by $\epsilon$
contain the families of Gaussian wavepackets having equal expected
energy. An energy conserved collapse model must map points from such a
curve onto some other point on the same curve. The curves in question
are given by

\begin{equation}
      \sigma^2 + (a-c)^2 + \frac{\hbar^2}{4mk\sigma^2} - \frac{2\epsilon}{k}=0.
      \label{eq:curvesofconstantE}
\end{equation}

It is seen that the resulting curves
(Fig.~\ref{fig:expectedEnergyconstraint}(b)) do not extend over the
entire domain of the mean position, $a$. Coupled with the fact that
Gaussian wavepackets are nonzero everywhere, this implies that an
energy conserved collapse to a Gaussian wavepacket in a harmonic
potential imposes a constraint on the locations where the
wavefunctions can collapse, even though the probability of finding
both particles at these points may be non-zero. In that sense, the
energy conserved collapse postulate does not strictly follow 
Born's rule.

The solution of equation (\ref{eq:curvesofconstantE}) produces the two possible values of the variance that a Gaussian wavepacket with expected energy $\epsilon$ and centered at $a$ must have
\begin{equation}
      \sigma^2 =  \frac{\epsilon}{k} + \frac{  (a-c)^2 \pm \sqrt{\left(\frac{2\epsilon}{k} - (a-c)^2\right)^2-\frac{\hbar^2}{mk}}}{2}
      \label{eq:variance}
\end{equation}

\subsubsection{Individual particles conserve energy}

\begin{figure*}
    \centering
    \includegraphics[width=0.28\linewidth]{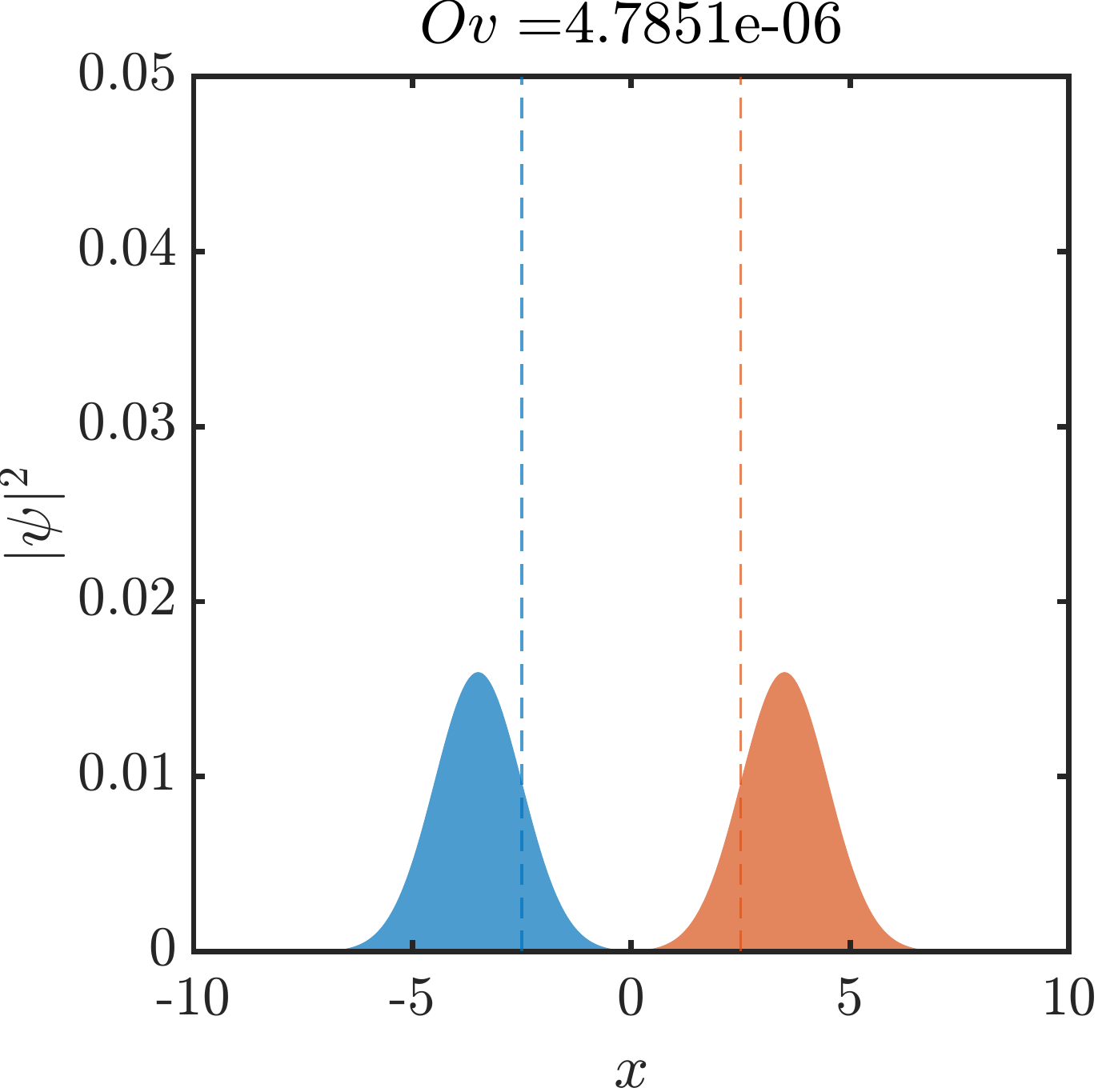}   \hspace{0.02\linewidth} \includegraphics[width=0.28\linewidth]{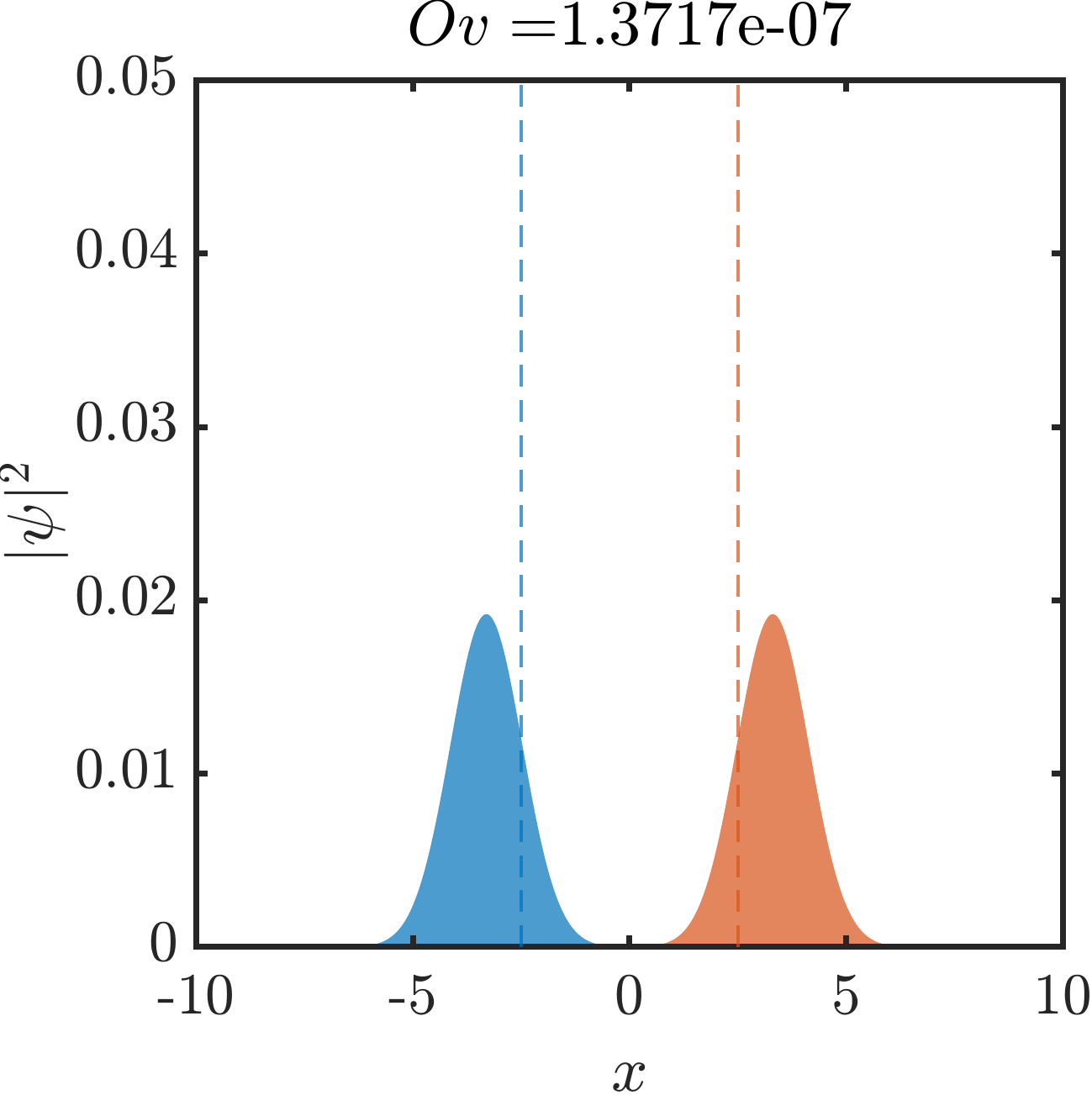}
    \hspace{0.02\linewidth}
    \includegraphics[width=0.28\linewidth]{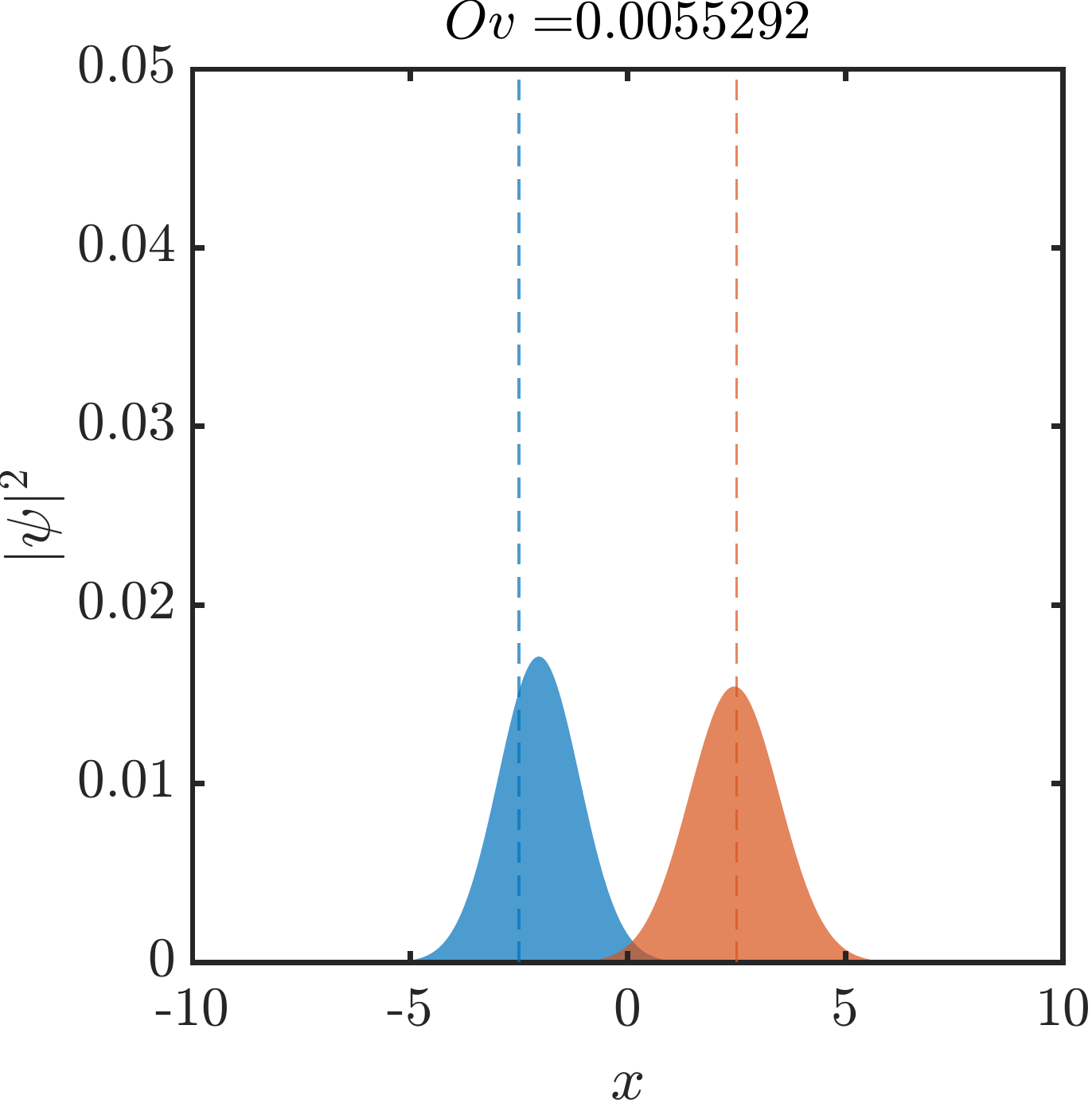}\\ \vspace{0.1in}
    \includegraphics[width=0.28\linewidth]{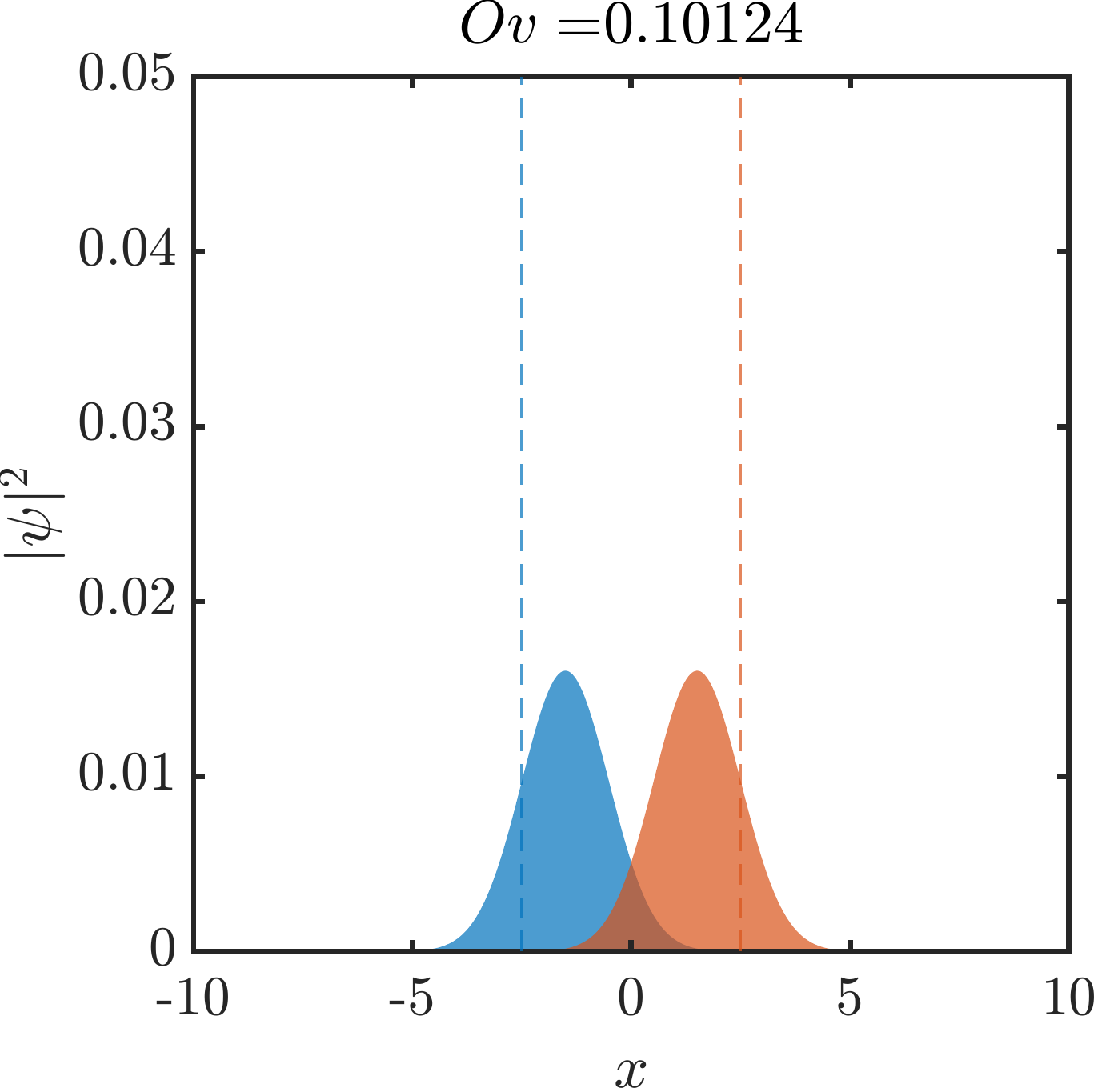}
    \hspace{0.02\linewidth}
    \includegraphics[width=0.28\linewidth]{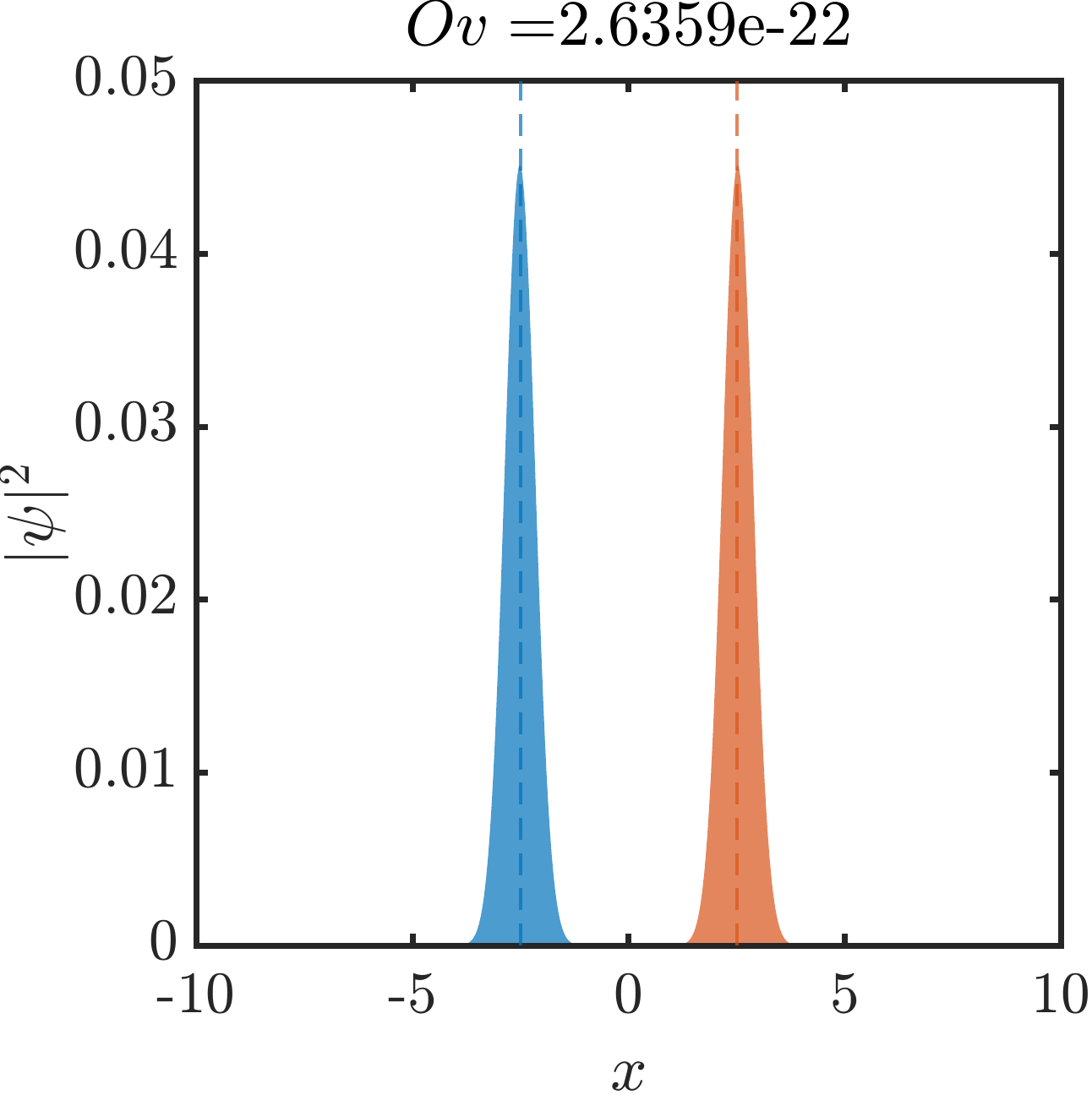}   \hspace{0.02\linewidth} \includegraphics[width=0.28\linewidth]{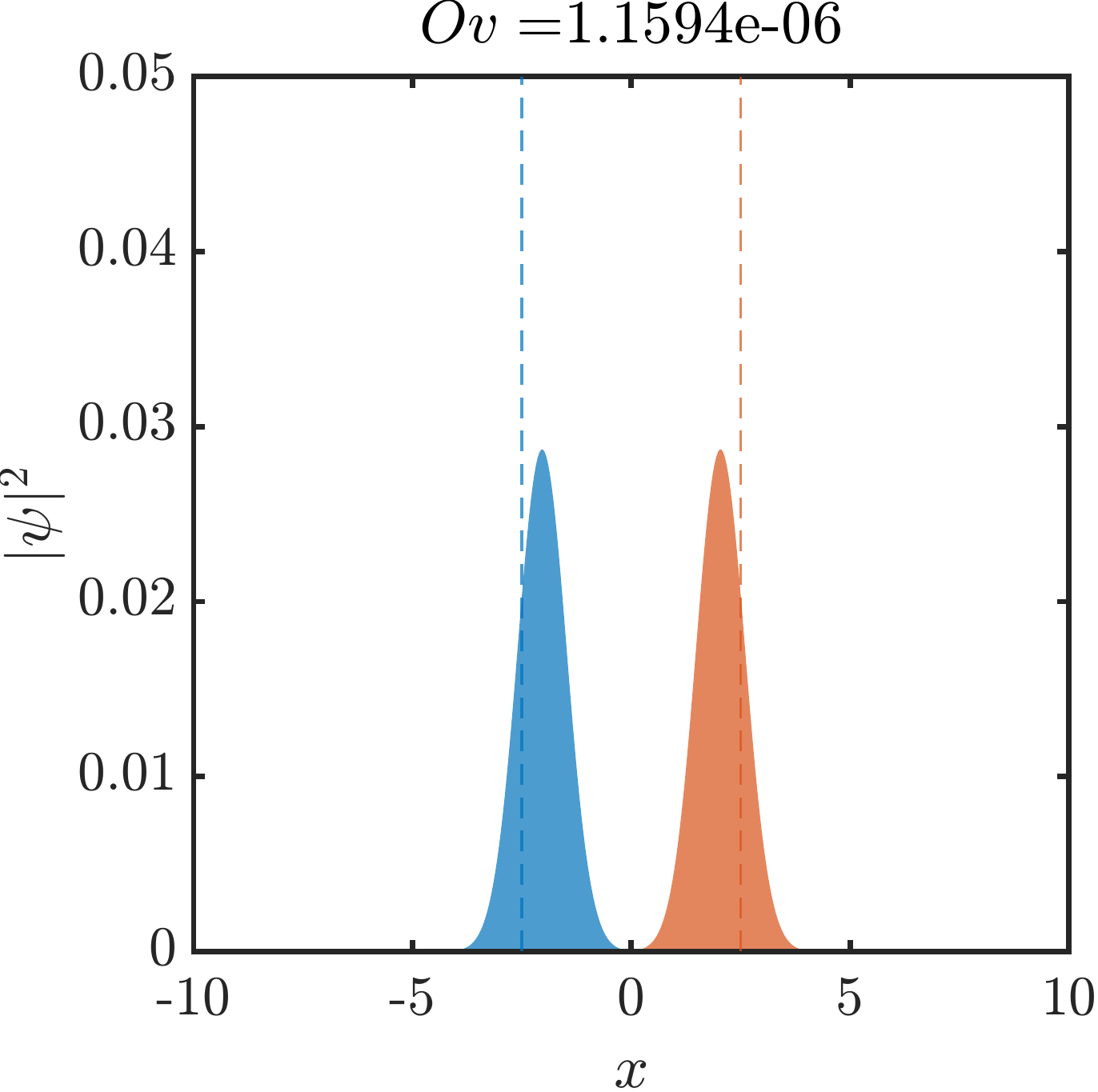}\\ \vspace{0.1in}
    
    \caption{ Time evolution of the probability distribution of the
      two particles at 6 time instants. The first and second particles
      are represented in blue and orange respectively, as are the
      centres of their respective harmonic potentials using dashed
      lines. Notice that the overlap between the distributions shown
      above the figures achieves a significant value in the fourth
      instant, triggering a collapse in the fifth instant. }
    \label{fig:evolve_individual}
\end{figure*}

The criteria we use for collapse is akin to the case of the soft
impact oscillator. At a time instant $t_i$, the probability of collapse
is the overlap between the densities of the two particles.
\begin{equation}P(t_i)=\int_{\infty}^{\infty} |\psi_1(x,t_i)|^2~ |\psi_2(x,t_i)|^2 ~\text{d}x
\end{equation}
For each particle to conserve energy individually, their
initial expected energy has to match their expected energy after
collapse. From Fig.~\ref{fig:expectedEnergyconstraint}(b) it is
apparent that in order to conserve energy, the wavefunction cannot
collapse at a position beyond the extent of the closed curve along
$a$. Within this limited domain, a collapse can follow Born's rule.

\begin{equation}
P(x) = \frac{{\lvert \psi(x) \rvert}^2}{ \displaystyle \int_{a_\text{min}}^{a_\text{max}}{\lvert \psi(x') \rvert}^2 ~\text{d}x' } ~, \; x \in [a_\text{min}, a_\text{max}]
\end{equation}
The collapse process is stochastic in both position and time, governed
by $P(x)$ and $P(t_i)$. When a collapse does occur, the wavefunctions
of both particles collapse simultaneously. For a particular position
of collapse $a$, the energy conservation constraint allows for two
different values of standard deviation $\sigma$ for the post collapse
wavepacket (see Fig.~\ref{fig:expectedEnergyconstraint}(b)). We choose the
smaller of the two to achieve maximal localization in position. Between
collapses, the two particles undergo unitary evolution according to
their respective Schr\"odinger equations. The resulting dynamics are
depicted in terms of a few frames at different time instants in
Fig.~\ref{fig:evolve_individual}.

\subsubsection{The system as a whole conserves energy}
Now we consider the case where each individual particle might lose or
gain energy but the expected energy of the two-particle system remains
conserved through the process of collapse, i.e.,
\begin{equation*}
\langle E \rangle_1+\langle E \rangle _2 = \langle E \rangle _\text{total}
=\text{constant.}
\end{equation*}
The conditions for collapse are calculated exactly as in the previous scenario. $P(t_i)$ and $P(x)$ dictate when and where the collapse happens. In this case, however, the energy must also be redistributed between the particles. 

\begin{figure}[tbh]
    \centering
    \includegraphics[width=0.85\linewidth]{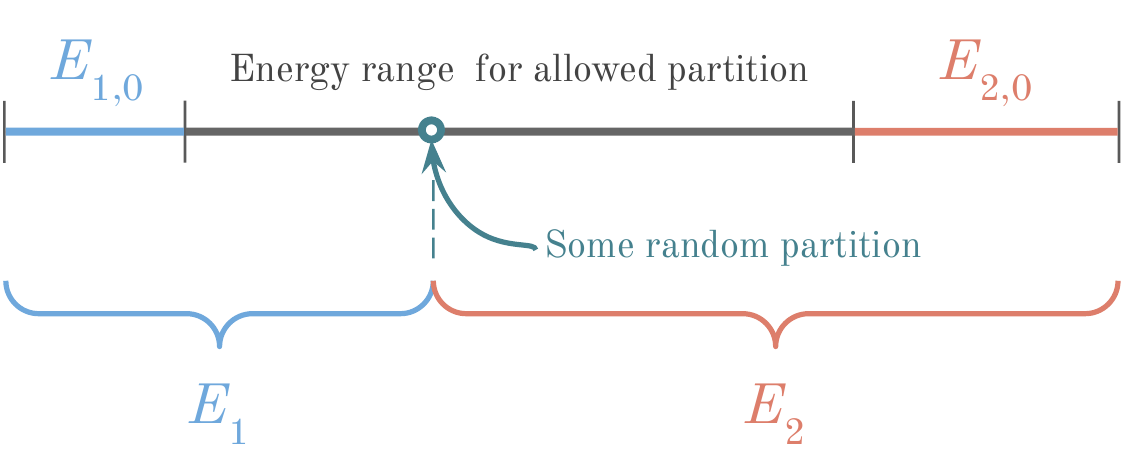}
    \caption{Schematic of the partitioning of energy between the two particles at the time of collapse. The length of the total line segment represents the total energy. The point of partition is chosen randomly on the allowed segment.}
    \label{fig:partition}
\end{figure}

 We chose to partition the total energy between the two particles randomly.  The minimum allowed energy for a particle in a harmonic oscillator potential is the ground state energy. Hence, an allowed partition must allot at least the respective ground state energy to each particle. This gives us an allowed range in energy for the partition. The partition is realized as a sample from a uniform distribution in the allowed energy range (Fig. \ref{fig:partition}).

\begin{figure*}
    \centering
    \includegraphics[width=\linewidth]{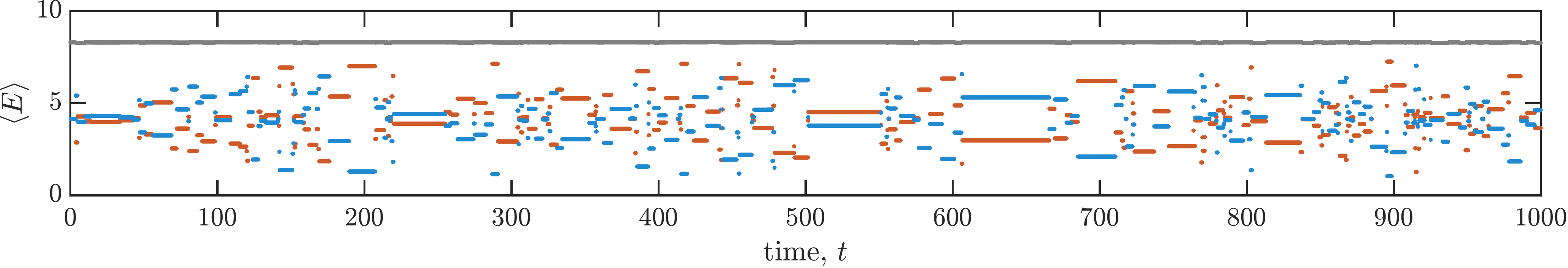}
    \caption{Expectation value of energy for the first particle (blue), second particle (orange) and the total system (black) for 1000 time steps. Energy is shuttled back and forth between the two particles, all the while conserving the total energy. }
    \label{fig:E1_E2}
\end{figure*}

Fig.~\ref{fig:E1_E2} shows that the expectation values of energy of
the particles vary in steps whenever collapses happen, while the total energy remains conserved. This behavior may be experimentally testable.

\begin{figure}[tbh]
    \centering
    \includegraphics[width=0.95\linewidth]{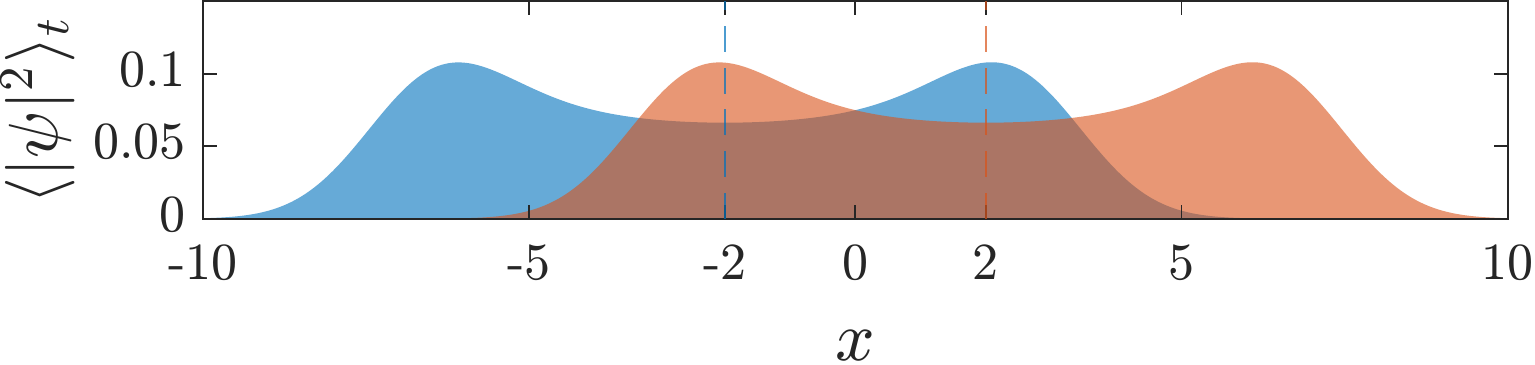} \\{\footnotesize (a)}\\ \vspace{0.2in}
    \includegraphics[width=0.95\linewidth]{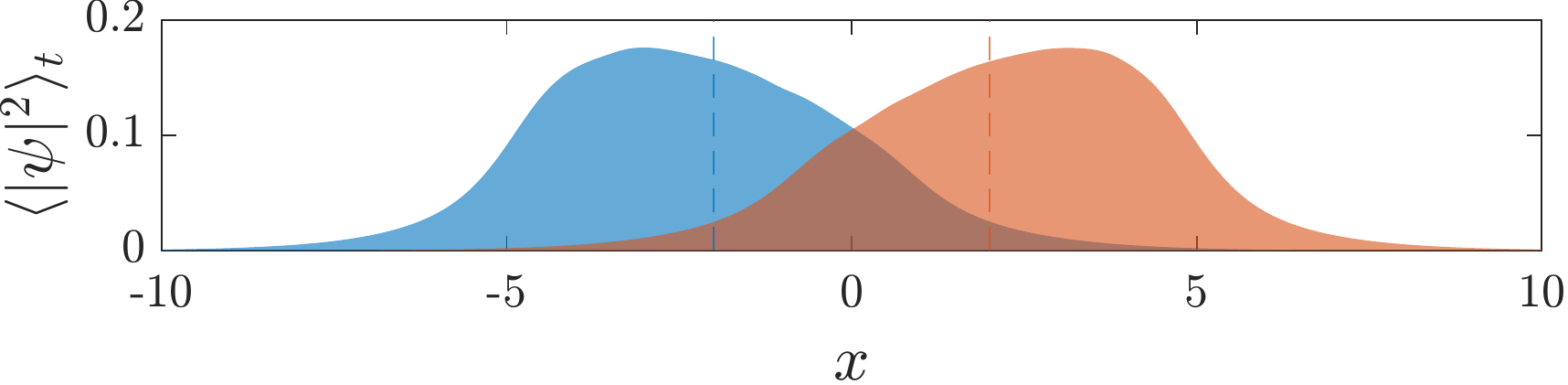}\\{\footnotesize (b)}\\ \vspace{0.2in}
    \includegraphics[width=0.95\linewidth]{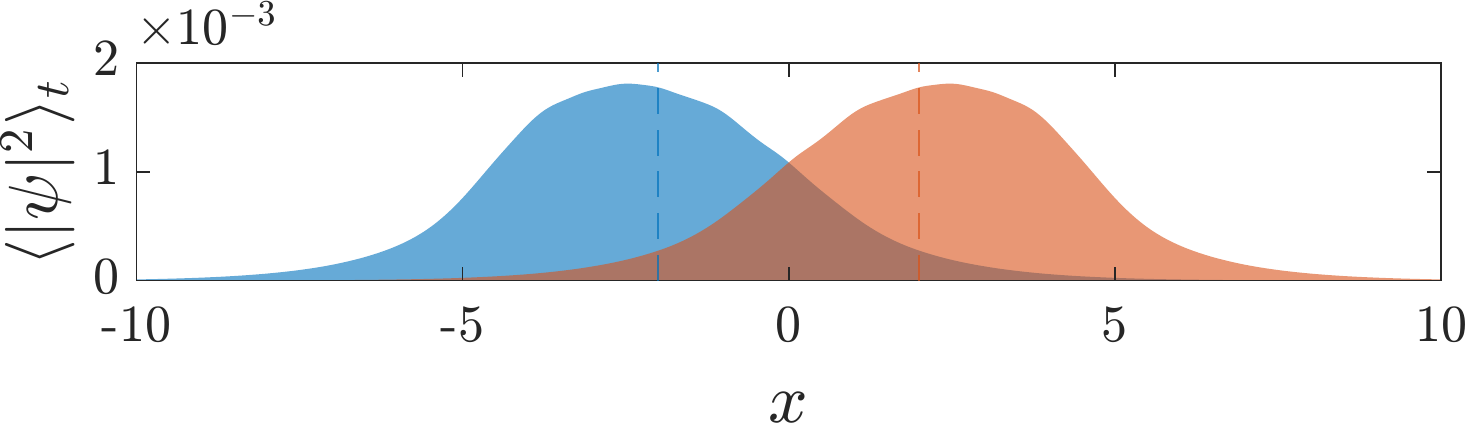}\\{\footnotesize (c)}
    \caption{The time-averaged probability density functions of position of the two particles for the cases of (a) no collapse, (b) collapse with individual particles conserving energy, (c) collapse with total system conserving energy. The particles are distinguishable with the parameters : $k_1=k_2=m_1=m_2=1$.  The centres of the respective potentials, $c_1=-2$ and $c_2=2$ are marked with dashed lines. }
    \label{fig:pos_dist_2P}
\end{figure}

\subsection{Comparison of the three cases}
To quantify the differences between the three cases considered, we plot the time-averaged probability density in position for the three cases in Fig.~\ref{fig:pos_dist_2P}. Without collapse, the average density functions have a two-humped form and are symmetric for both particles. For both the collapse postulates, the densities have single peaks and are very similar. For the case where the individual energies are conserved, the distribution functions are more skewed
(i.e., the probability of finding the particles away from each other is higher) than the case where the total energy is conserved. An observed increase in the average distance between the particles can serve as a testable confirmation of the postulated collapse processes. The amount of increase can narrow down the type of interaction. 

\begin{figure} 
  \centering
    \includegraphics[width=0.95\linewidth]{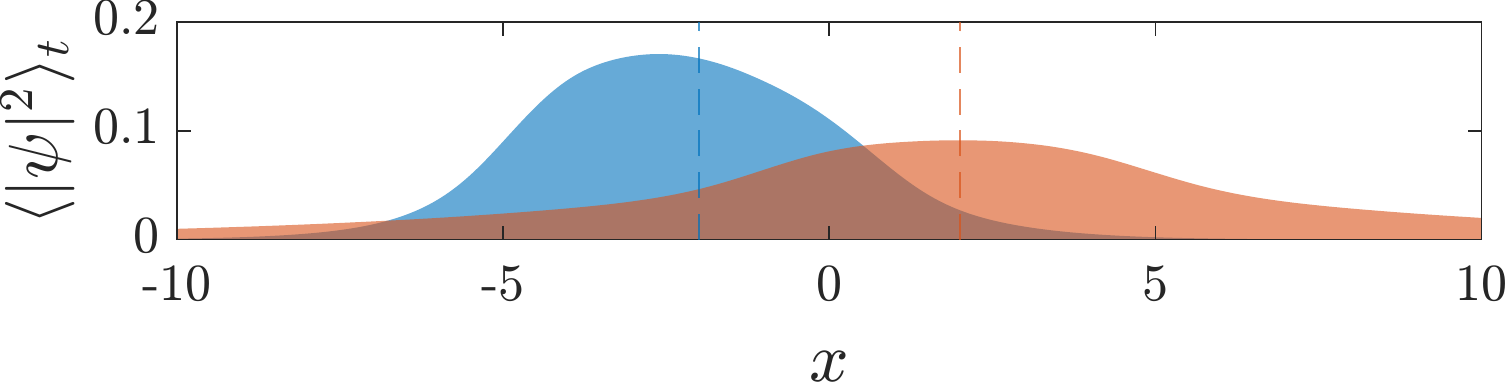}\\\hspace{0.08\linewidth}{\footnotesize (a)}\\
    \vspace{0.2in}
    \includegraphics[width=0.95\linewidth]{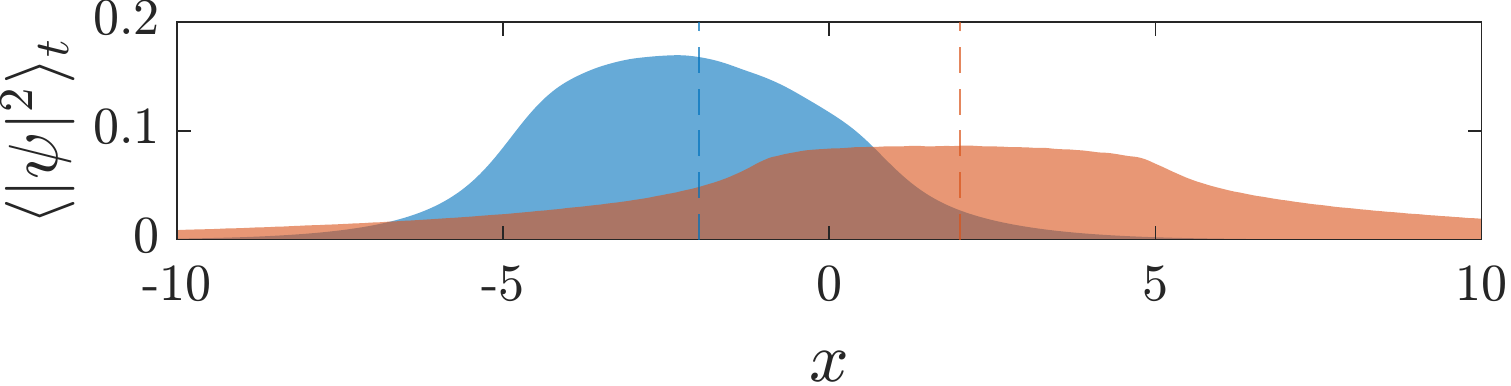}\\\hspace{0.08\linewidth}{\footnotesize (b)}
\caption{The energy conserved collapse scenario when the two particles
  have dissimilar parameters: $k_1=m_1=1$, $k_2=10$ and $m_2 =
  0.1$. (a) individual particles conserving energy and (b) total system
  conserving energy.}\label{fig:pos_dist_3P}
\end{figure}

Fig.~\ref{fig:pos_dist_3P} shows the case when the $m$
and $k$ values of the two oscillators are different. In this case the qualitative behavior is similar, but the density function of the
particle of lower mass has an almost flat top.

\section{Conclusion}

In this work we have formulated a few alternate postulates for the collapse of the wavefunction. We assume that collapse of the wavefunction of a particle does not depend on the intervention of a conscious observer. Instead, its interaction with another object, classical or quantum, may collapse its wavefunction. The probability of the occurrence of such a collapse depends on the overlap between
the wavefunctions of the interacting entities.

For the situation where the particle interacts with a classical
object, we have formulated four different postulates regarding the
condition of collapse and the post-collapse wavefunction. For
interaction among two distinguishable quantum particles, we have
formulated two postulates of energy conserving wavefunction collapse.

We have proposed a model system---the quantum version of a soft-impact
oscillator---and have obtained testable predictions from each
postulate regarding the energy and position distribution in such a
system. Our computations predict that, if an interaction with a
classical object induces collapse of the wavefunction, then the
probability distributions of energy and position would be different
from what is predicted by standard quantum mechanics. If an
interaction between two distinguishable quantum particles can induce
energy-conserving collapse of their wavefunctions, then the average
distance between them would tend to be larger than the distance
between the centres of the two potential functions.
Experimental test of the predictions would enable us to
eliminate the wrong postulates.

Notice that, if any of the postulates regarding collapse of the
wavefunction is supported by experiment, it will have important
consequence in the foundation of quantum mechanics. It will imply that
collapse of a wavefunction is a natural process that does not require
conscious observation.


\end{document}